\documentclass[10pt,twocolumn,letterpaper]{article}
\pdfoutput=1

\usepackage{iccv}
\usepackage{times}
\usepackage{epsfig}
\usepackage{graphicx}
\usepackage{amsmath}
\usepackage{amssymb}
\usepackage{xcolor}
\usepackage{soul}
\usepackage{subcaption}
\usepackage{enumitem}

\usepackage[pagebackref=true,breaklinks=true,letterpaper=true,colorlinks,bookmarks=false]{hyperref}

\iccvfinalcopy 

\graphicspath{{./figs/}{./images/}}

\DeclareMathOperator*{\argmin}{arg\,min}

\newcommand{\norm}[1]{\left\lVert#1\right\rVert}

\def\iccvPaperID{1} 

\ificcvfinal\fi

\begin{document}

\title{Temporal Kernel Consistency for Blind Video Super-Resolution}


\author{Lichuan Xiang$^{1*}$, Royson Lee$^{2}\thanks{Equal contributions.}$, Mohamed S. Abdelfattah$^3$,  \\  Nicholas D. Lane$^{2,3}$, Hongkai Wen$^{1,3}$ \\
{\normalsize $^1$University of Warwick \phantom{0} $^2$University of Cambridge \phantom{0} $^3$Samsung AI Center, Cambridge} \\
{\tt\footnotesize l.xiang.2@warwick.ac.uk}}


\maketitle
\ificcvfinal\thispagestyle{empty}\fi

\begin{abstract}
    Deep learning-based blind super-resolution (SR) methods have recently achieved unprecedented performance in upscaling frames with unknown degradation. These models are able to accurately estimate the unknown downscaling kernel from a given low-resolution (LR) image in order to leverage the kernel during restoration. Although these approaches have largely been successful, they are predominantly image-based and therefore do not exploit the temporal properties of the kernels across multiple video frames. In this paper, we investigated the temporal properties of the kernels and highlighted its importance in the task of blind video super-resolution. Specifically, we measured the \textit{kernel temporal consistency} of real-world videos and illustrated how the estimated kernels might change per frame in videos of varying dynamicity of the scene and its objects. With this new insight, we revisited previous popular video SR approaches, and showed that previous assumptions of using a fixed kernel throughout the restoration process can lead to visual artifacts when upscaling real-world videos. In order to counteract this, we tailored existing single-image and video SR techniques to leverage kernel consistency during both kernel estimation and video upscaling processes. Extensive experiments on synthetic and real-world videos show substantial restoration gains quantitatively and qualitatively, achieving the new state-of-the-art in blind video SR and underlining the potential of exploiting kernel temporal consistency. 
\end{abstract}

\section{Introduction} 
\label{sec:intro}

Super-resolution (SR) is an ill-posed problem that assumes the low-resolution image (LR) is derived from a high-resolution (HR) image and is recently dominated by deep learning due to its unprecedented performance~\cite{SRCNN}.
In order to better restore the high-frequency details, state-of-the-art video SR methods~\cite{TDAN, EDVR, Xiang2020} exploit the temporal frame information by employing a multi-frame SR (MFSR) approach.
Specifically, each supporting frame is aligned with its reference frame through motion compensation before the information in these frames are merged for upscaling.

Most of these methods, however, assume that the degradation process, applying the blur kernel and the downscaling operation, is pre-defined.
Therefore, the performance of these methods significantly deteriorates for real-world videos as the downscaling kernel, which is used for upscaling, differs from the ground truth kernel, a phenomenon known as the kernel mismatch problem~\cite{Efrat2013}.
Although there has been significant progress to enable the usage of SR models in real-world applications, these solutions are predominantly image-based~\cite{KernelGAN,IKC,DAN,KMSR}.
The primary paradigm of these blind image-based solutions consists of either a two-step or an end-to-end process, starting with a kernel estimation module and followed by a SR model that aims to maximize image quality given the estimated kernel and/or noise. 
Hence, when upscaling videos, these works do not utilize the temporal similarity between kernels and have to estimate kernels individually per frame. 
This is not only computationally expensive but also less effective, since estimating kernels independently per-frame may result in inaccurate kernels, as shown in Sec.~\ref{sec:tke} and Sec.~\ref{sec:importance_tke}, and thus kernel mismatch. 

Recent blind MFSR approaches, on the other hand,  utilized a fixed kernel to upscale every frame~\cite{Liu2014,Pan2020} in the same video -- we hypothesize that this fixed kernel assumption can also lead to kernel mismatch.
Therefore, in this work, we attempt to investigate and answer the following questions: \textit{how does the kernel change temporally in real-world videos, and how can we leverage this change in the video restoration process?}

Towards our goal, we first investigated the temporal differences in kernels in Sec.~\ref{sec:tke}. 
In particular, we used a recent image-based kernel estimation approach, KernelGAN~\cite{KernelGAN}, on frames of real-world videos and observed that videos of varying dynamicity, such as scene changes and object motion blurs, can result in corresponding variations in the downsampling kernels. 
We then show how videos of different dynamicity can affect the temporal consistency of their downscaling kernels.
From this perspective, we re-evaluated previous MFSR approaches on real-world videos in Sec.~\ref{sec:importance_tke}.
Through our experiments, we show that the common assumption of using a fixed downsampling kernel for multi-frame approaches can lead to the kernel mismatch problem, resulting in inaccurate motion compensation and hence inferior restoration results.
To counteract these drawbacks, we tailored these existing techniques to exploit our new insight on \textit{kernel temporal consistency} in Sec.~\ref{sec:utilizing_tke}, leading to substantial gains as compared to state-of-the-art.
In summary, the main contributions of this work are:
\begin{itemize}
\setlength\itemsep{0mm}
\item To the best of our knowledge, we are the first to investigate the temporal consistency of kernels in real-world videos for deep blind video SR. 
\item We present the limitations and drawbacks of using a fixed kernel, a scenario that is commonly assumed, for multi-frame SR approaches.
\item Through tailored alterations to existing SR approaches, we underline the potential of exploiting \textit{kernel temporal consistency} for accurate kernel estimation and motion compensation, resulting in considerable performance gains in video restoration.

\end{itemize}

\section{Related Work} 
\label{sec:related_work}

\noindent \textbf{Single-Image Blind Super-Resolution. }
Previous deep learning-based image-based SR approaches~\cite{SRCNN, FSRCNN, ESPCN, TPSR, ESRN, CARN, ESPCN} assumed a fixed and ideal downsampling degradation process, often bicubic interpolation, leading to poor performance when applied to real-world images.
As a result, most blind SR approaches focused on estimating the downsampling kernel and/or utilizing it for upsampling.
Efrat \textit{et al.}~\cite{Efrat2013} first highlighted the kernel mismatch problem: using the incorrect kernel during restoration had a significant impact on the performance regardless of the choice of image prior.
Towards accurate downsampling kernel estimation, Michaeli \textit{et al.}~\cite{Michaeli2013} exploited the inherent recurrence property of image patches and proposed an iterative algorithm to derive the kernel that maximizes the similarity of recurring patches across scales of the LR image.
Bell-Kligler \textit{et al.}~\cite{KernelGAN} adopted a GAN approach~\cite{Goodfellow2014}, in which the generator learnt an estimated kernel to downscale the input image with and the discriminator learnt to differentiate between the patch distribution of the input image and its downscaled variant.
The downsampling kernel can also be learned using CNNs by enforcing that the super-resolved image maps back to the LR image~\cite{DRN} or using a paired of real-world image dataset~\cite{Cai2019}.
Exploiting the kernel mismatch phenomenon, Gu \textit{et al.}~\cite{IKC} and Luo \textit{et al.}~\cite{DAN} alternatively estimated the kernel from the approximated super-resolved image and restored the image by using the estimated kernel, reaching the current state-of-the-art.

\noindent \textbf{Multi-Frame Super-Resolution.}
MFSR approaches focus on utilizing temporal information from the LR frames by aligning and fusing them in order to further boost restoration performance through CNNs or RNNs. 
Earlier works~\cite{Liao2015, Kappeler2016} performed motion compensation by estimating optical flow using traditional off-the-shelf motion estimation algorithms~\cite{Baker2007}.
As the accuracy of motion estimation directly affects the reconstruction quality of the super-resolved images, these traditional motion estimation works are superseded by more accurate CNN-based networks such as spatial transformer networks~\cite{Jaderberg2015} or task-specific motion estimation networks~\cite{FlowNet2, SPynet, PWCNet}, leading to approaches~\cite{Liu2017,Makansi2017,Tao2017,Xue2018,FRVSR} that focused on integrating motion estimation and SR networks for end-to-end learning.
Recent works~\cite{Jo2018, TDAN, EDVR, Xiang2020} decoupled this dependency on motion estimation networks and performed motion compensation by adaptively aligning the reference and supporting frames through dynamically-generated filters or deformable convolutions~\cite{Dai2017, Zhu2019}.
Although majority of these works helped to elucidate the relationship between motion estimation and video restoration, they neglected the degradation process by assuming a fixed known kernel.
Therefore, unlike previous MFSR works that focused on incorporating temporal information in the frames, we also utilized the temporal information in the downscaling degradation operation in order to further boost restoration performance. 

Towards blind MFSR, Pan \textit{et al.}~\cite{Pan2020} used a kernel estimation network, consisting of two fully-connected layers, to learn a fixed blur kernel for inference.
However, they, similar to Liu \textit{et al.}~\cite{Liu2014}, assumed that the kernel is fixed at every timestamp, resulting in poor SR performance as shown in Sec.~\ref{sec:importance_tke}. 



\section{Problem Formulation}\label{sec:prob_formulation}

Multi-frame Super Resolution (MFSR) uses a set of $2N$ supporting LR frames $\{y_{t-N}, \dotsb, y_{t-1}, y_{t+1}, \dotsb, y_{t+N}\}$ to upscale the reference LR frame $y_t$ at time $t$, utilizing temporal information across frames. The degradation process is usually expressed as follows:
\begin{equation}\label{eq:mfsr}
    y_{t+i} = \big((F_{t\rightarrow{t+i}}x_t) * k_{t+i}\big)\downarrow_s + n_{t+i}
\end{equation}
where $y$ and $x$ are the LR and HR image respectively, $k$ is the blur kernel, $\downarrow_s$ is the downscaling operation (\textit{e.g.} sub-sampling) using scaling factor $s$, $n$ is the additive noise, $i={-N, \dotsb, N}$, and $F$ is the warping matrix w.r.t the optical flow applied on $x_t$. The image warping process can either be done explicitly via an optical flow or implicitly via dynamically-generated filters~\cite{Jo2018} or deformable convolutions~\cite{Dai2017}.
The process of applying $k$ together with the $\downarrow_s$ is also referred to as applying a downscaling kernel or SR kernel~\cite{KernelGAN, IKC}. 
Traditionally, a prior term is individually hand-crafted for $x_t$, $k_{t+i}$ and ${F_{t\rightarrow{t+i}}}$, but most deep learning-based approaches capture the prior~\cite{SRCNN} through CNNs by training it using a large amount of examples. 

In order to solve for $k$ and $x$, state-of-the-art blind image-based algorithms split the problem into two sub-problems, estimating $k$ and restoring $x$, and address each problem sequentially~\cite{KernelGAN, IKC} or alternately~\cite{USRNet, DAN}.
MFSR solutions, on the other hand, include an additional sub-problem of estimating the motion between each supporting frame and its reference frame in order to perform motion compensation and hence leverage the temporal frame information during restoration.
Although previous traditional video SR approaches~\cite{Farsiu2004,Ma2015} assume that the kernel varies across frames, recent works~\cite{Liu2014,Pan2020} assume a fixed kernel.
In our work, we study and highlight the implications of both assumptions and advocate for the per-frame kernel approach, resulting in the following optimization problem:
\begin{equation} \label{eq:mfsr_opt}
\begin{split}
    \hat{x}_t & = \argmin_{x_t}\sum^{N}_{i={-N}}\norm{y_{t+i} - \big((F_{t\rightarrow{t+i}}x_t) * k_{t+i}\big)\downarrow_{s}}\\
    \hat{k}_t & = \argmin_{k_t} \norm{y_{t} - (x_t * k_{t})\downarrow_s}\\
    \hat{F}_{t\rightarrow{t+i}} & = \argmin_{F_{t\rightarrow{t+i}}}\norm{y_{t+i} - \big((F_{t\rightarrow{t+i}}x_t) * k_{t+i}\big)\downarrow_{s}}
\end{split}
\end{equation}
where $\hat{x}$, $\hat{k}$, and $\hat{F}$ are the estimated HR image $x$, kernel $k$, and warping matrix $F$ respectively. 

\section{Kernels In Real-World Videos}\label{sec:tke}


\begin{figure}[t]
 \centering
 \includegraphics[width=\columnwidth]{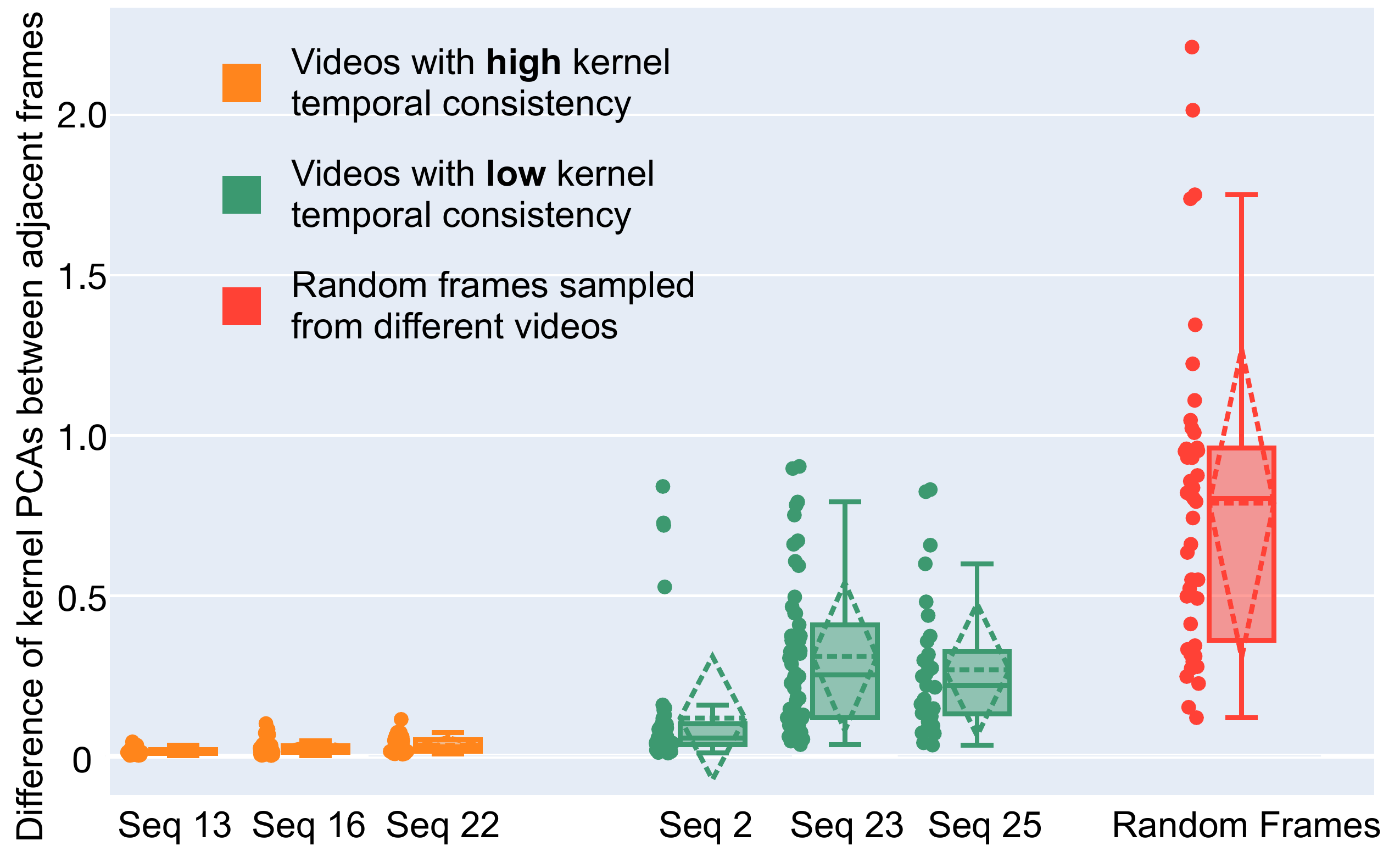}
 \caption{We quantify temporal kernel consistency by measuring kernel PCA change for adjacent frames in real-world videos with high/low kernel temporal consistency. Random frames are sampled from different videos at each timestamp as a baseline to highlight temporal kernel consistency within same video. Kernel changes are represented by solid dots while boxplots show distributions. }
 \label{fig:consistency_comparison}
\end{figure}

\begin{figure*}[t]
 \centering
   \includegraphics[width=\textwidth]{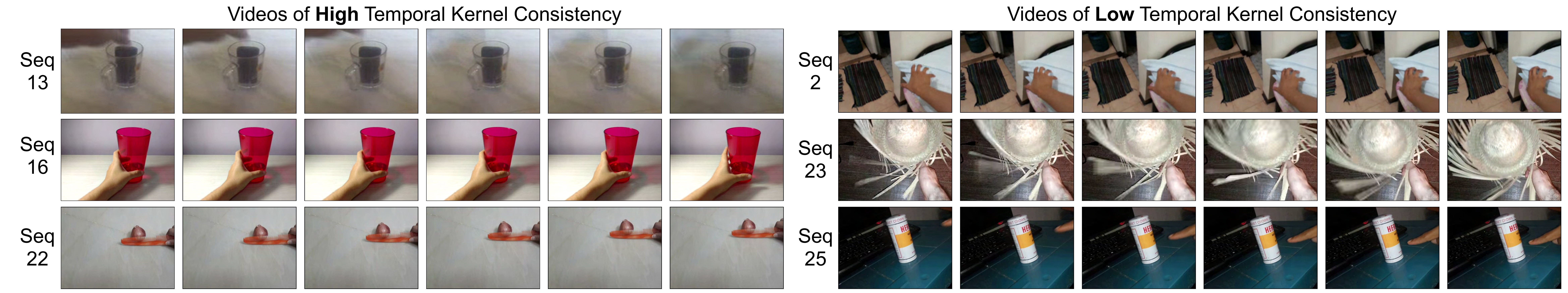}
 \caption{Example frames from videos of high/low kernel temporal consistency as shown in Fig.~\ref{fig:consistency_comparison}. Videos of low kernel temporal consistency (right) contain a higher proportion of video dynamicity as compared to videos of high kernel temporal consistency (left).} 
 \label{fig:vid_eg}
\end{figure*}

In order to investigate the temporal kernel changes in real-world videos, we extracted a pool of kernel sequences from the Something-Something dataset~\cite{something}, a real-world video prediction dataset. 
As ground truth kernels do not exist in real-world videos, we applied the state-of-the-art image-based kernel extraction method, KernelGAN~\cite{KernelGAN}, to extract the sequences of kernels.
Through these kernel sequences, we observed that the extracted SR kernels can often be different for each frame, while on the other hand may also exhibit certain levels of temporal consistency, depending on the video's dynamicity.

Fig.~\ref{fig:consistency_comparison} illustrates this phenomena, in which we show the distributions of the magnitude of kernel changes in different video sequences. 
Specifically, we reshaped the extracted kernels for each frame and reduced them through principal component analysis (PCA). 
We then computed the sum of absolute differences between the kernel PCA components of adjacent frames and plotted this difference using videos of varying dynamicity (left and middle plot groups in Fig.~\ref{fig:consistency_comparison}).
As a baseline, in comparison with an unrealistic real-world video without any temporal consistency, we sampled random frames from different random videos at each timestamp, of which its kernel PCA changes are represented by the right plot in Fig.~\ref{fig:consistency_comparison}.
We observed that some videos' kernel differences, namely the left group of plots showing video sequences 13, 16 and 22 from the Something-Something dataset in Fig.~\ref{fig:consistency_comparison}, are of \textit{high temporal kernel consistency}, of which the kernels remain largely unchanged throughout. 
In contrast, the middle group of plots represent the kernel differences of videos with \textit{low temporal kernel consistency}, namely video sequence 2, 23 and 25, of which kernel changes can be much significant.
Visually, Fig.~\ref{fig:vid_eg} shows example frames from corresponding videos of high and low temporal kernel consistency.
In particular, videos with high temporal kernel consistency depict slow and steady movements with no motion blurs or scene changes - \textit{e.g.} a video of a hand slowly reaching towards a cup or videos with almost identical frames at each time step. 
On the other hand, videos with low temporal kernel consistency have motion blur caused by rapid movements of the camera or object, \textit{e.g.} large object motions caused by a man weaving a straw hat or placing a container upright and shaky camera motions, as illustrated in the right of Fig~\ref{fig:vid_eg}.
Therefore, our experiments highlight that SR kernels in real-world videos are often non-uniform and can exhibit different levels of temporal consistency.

\section{Kernel Mismatch in Previous MFSR }
\label{sec:importance_tke}

In order to highlight the importance of incorporating temporal kernel consistency in blind video restoration, we looked into the limitations and drawbacks of both previous and recent multi-frame super resolution (MFSR) approaches~\cite{Liao2015, Kappeler2016, Liu2017,Makansi2017,Tao2017,Xue2018, Pan2020}.
Specifically, these works assumed that either a fixed degradation operation is used for all videos or a fixed SR kernel is used to degrade all frames in each video -- assumptions that do not hold for real-world videos as shown in Sec.~\ref{sec:tke}.
Consequently, these works suffer from the kernel mismatch phenomenon~\cite{Efrat2013} when they are used to upscale real-world videos.

\begin{figure}[t]
 \centering
 \includegraphics[width=\columnwidth]{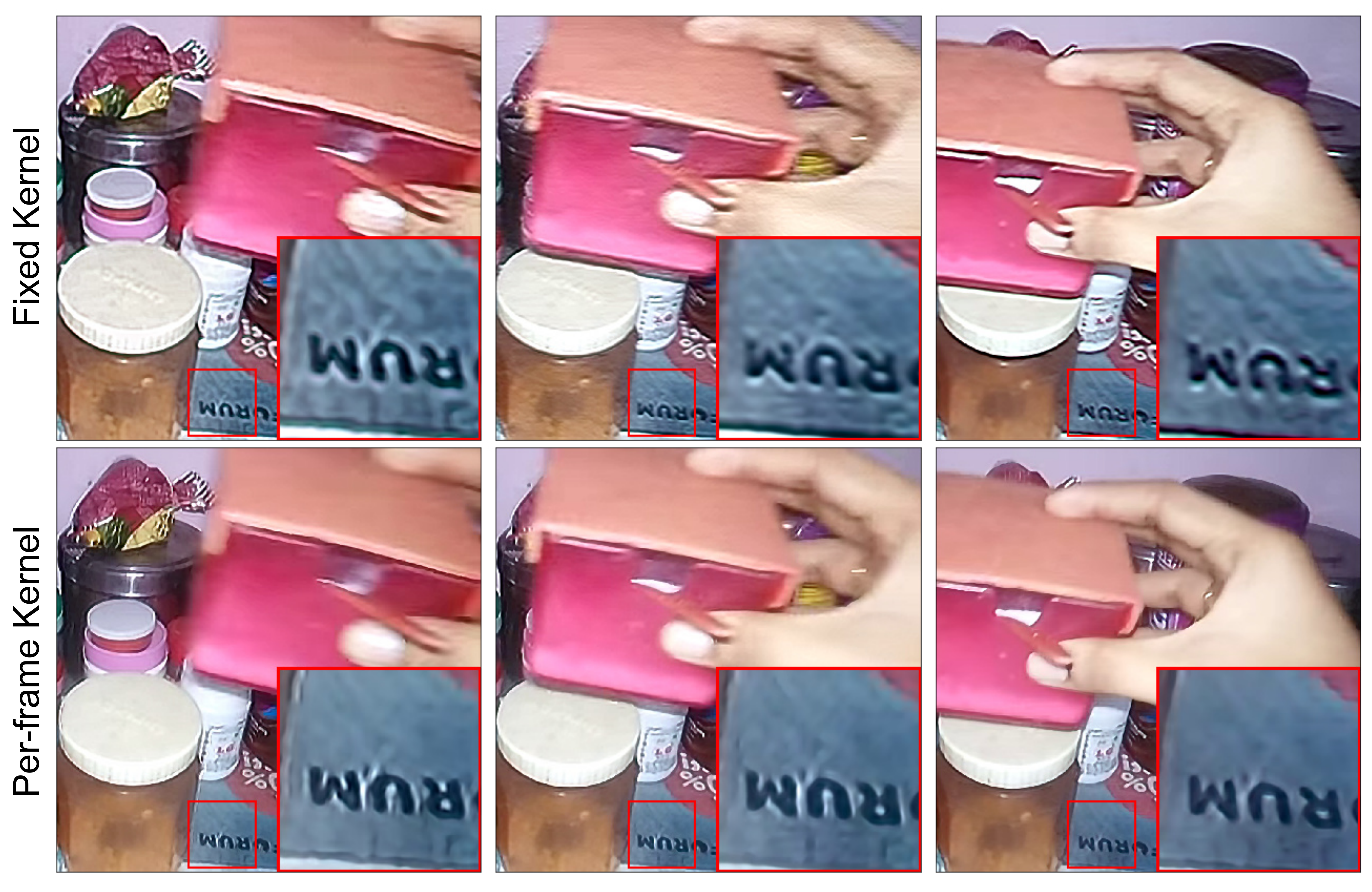}
 \caption{Examples of consecutive frames in real-world videos upscaled using a fixed kernel (top row), and different per-frame kernels (bottom row). More examples can be found in the supplementary material (s.m. Fig.~1).}
 \label{fig:fixed_kernel}
 \vspace{-3mm}
\end{figure}

\begin{figure*}[t]
 \centering
 \includegraphics[width=\textwidth]{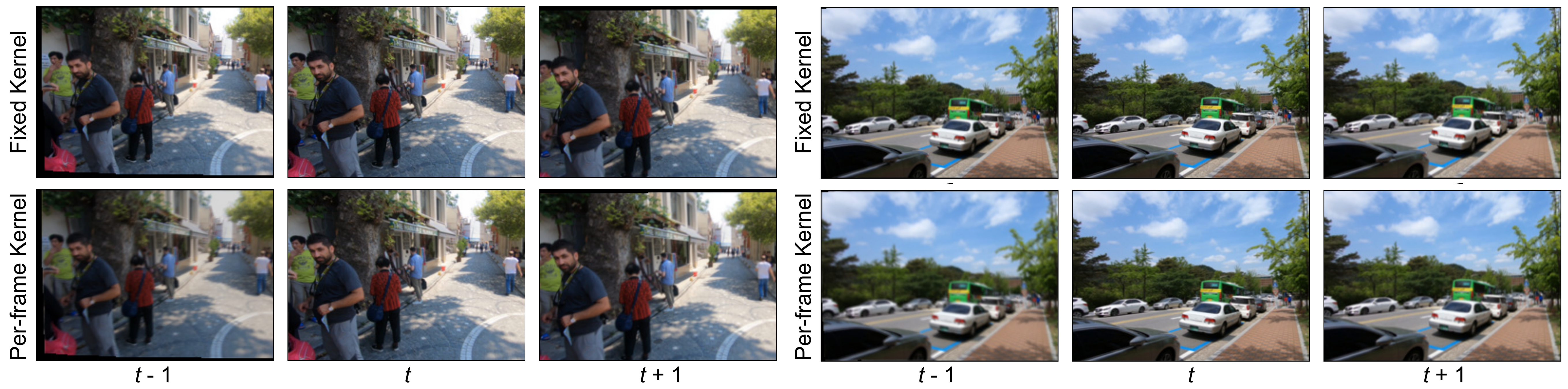}
 \caption{Example aligned frames with their reference frame at time $t$. Motion compensation model in current MFSR approaches performs better when considering a fixed SR kernel at every timestamp (top row), which however does not hold for real-world videos. For videos with varying kernels per frame, the aligned frames are oversmoothed and blurred (\textit{e.g.} see the frames at $t$ - 1 of both examples) due to kernel mismatch (bottom row). Zoom in for best results. More examples are provided in the supplementary material (s.m. Fig.~4.)}
 \label{fig:vid_flow}
 \vspace{-3mm}
\end{figure*}
\noindent \textbf{Impact on Frame Upscaling.}
Previous MFSR works exploit temporal frame information using a fixed SR kernel.
We first show with a naive approach, that the use of a single kernel, even \textit{without} utilizing temporal frame information, to restore every frame is detrimental towards the performance of frame upscaling in the video restoration process. 
Towards this goal, we independently computed a kernel per frame of videos taken from the Something-Something dataset using KernelGAN and restored these frames using ZSSR~\cite{ZSSR}. 
We compared this per-kernel approach with a single-kernel approach where we only estimated the SR kernel on the first frame in the video and used that same kernel to restore all subsequent frames in each video.
Fig.~\ref{fig:fixed_kernel} shows the qualitative difference between the two experiments and we observed that using a fixed kernel indeed resulted in more severe visual artifacts and unnatural textures. 
All experiment details and more examples can be found in the supplementary material (s.m. Sec.~1 \& Fig.~3).

\noindent \textbf{Impact on Motion Compensation.}
We then show that a fixed kernel assumption further aggravates MFSR approaches.
The premise of these approaches is to utilize temporal frame information in order to boost the restoration performance. 
To this end, previous MFSR works used motion compensation to warp each supporting frame to its reference frame before fusing these frames together for upscaling.
As mentioned in Sec.~\ref{sec:prob_formulation}, the optical flow used for warping is either estimated explicitly using traditional or deep motion-estimation techniques or implicitly using adaptive filters or deformable convolutions.

In order to visualize the impact of kernel mismatch on motion compensation for real-world videos, we consider two sets of videos, one from LR sequences of the original REDS dataset~\cite{REDS}, which are degraded using a fixed kernel, while the other from our \textit{REDS10} testing sequence (details discussed in Sec.~\ref{sec:setup}), which are generated using different per-frame kernels and thus better resemble the degradation characteristics of real-world videos than the former.

We then used an explicit deep motion estimation model, which is commonly used in previous MFSR approaches~\cite{Liu2017,Makansi2017,Tao2017,Xue2018} to compute the optical flow. Specifically, we adopt PWCNet~\cite{PWCNet} to estimate optical flow in our experiment.
The optical flow is then used to warp each supporting frame and the results are shown in Fig.~\ref{fig:vid_flow}, for both fixed and per-frame degradation video sets.
We observe that motion compensation performs better on the fixed degradation video set, benefiting the previous approaches that were specifically designed under the fixed kernel assumption.
On the other hand, due to the kernel dynamicity in real-world videos, the warped supporting frames of those approaches often suffer from kernel mismatch when dealing with videos of varying kernels, as shown in Fig.~\ref{fig:vid_flow} (bottom row).
We further show that this phenomenon is also observed with the use of implicit motion compensation, and the errors incurred from inaccurate motion compensation can be propagated throughout the restoration process in Sec.~\ref{sec:tke_effectiveness}.

\section{Exploiting Temporal Kernel Consistency}
\label{sec:utilizing_tke}

We hypothesize that by using temporal kernel consistency, we can mitigate the limitations highlighted in Sec.~\ref{sec:importance_tke}.
Towards understanding the impact of doing so, we first adopted the state-of-the-art blind image-based SR algorithm, DAN~\cite{DAN} and incorporated MFSR modules from EDVR~\cite{EDVR} for temporal alignment through implicit motion compensation, fusion, and video restoration.
We then tailored these approaches to exploit temporal kernel consistency and analyzed the benefits and performance impact of doing so through an ablation study. 

\subsection{Experiment Setup}\label{sec:setup}

\begin{figure}[t]
 \centering
 \includegraphics[width=\columnwidth]{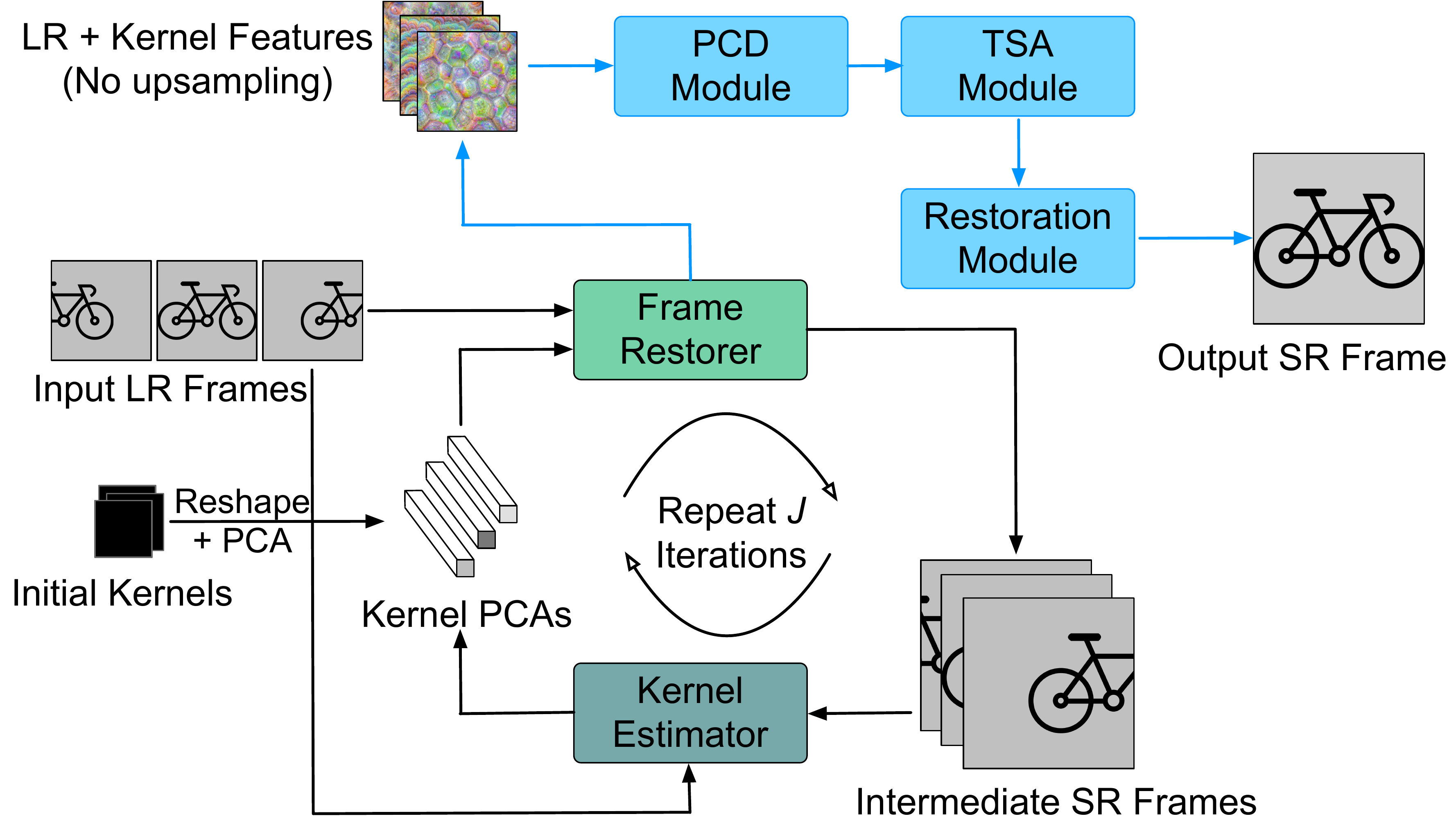}
 \caption{Our experiment setup of utilizing multiple frames for temporal kernel estimation (shown in black) and using temporal kernels for multi-frame restoration (shown in blue). See text for details and the supplementary material (s.m. Fig.~1) for a more detailed architecture diagram. }
 \label{fig:overview}
 \vspace{-4mm}
\end{figure}

\noindent \textbf{Models.}
DAN~\cite{DAN} is an end-to-end learning approach that estimates the kernel $k$, and restores the image $x$, alternately.
The key idea, as shown in black in Fig.~\ref{fig:overview}, is to have two convolutional modules: 1) a \textit{restorer} that reconstructs $x$ given the LR image $y$, and the PCA of $k$; and 2) an \textit{estimator} that learns the PCA of $k$, based on $y$ and the resulting super-resolved image $\hat{x}$.
The basic block for both components is the conditional residual block (CRB), which concatenates the basic and conditional inputs channel-wise and then exploit the inter-dependencies among feature maps through a channel attention layer~\cite{RCAN}. 
The alternating algorithm executes both components iteratively, starting with an initial kernel, Dirac, and resulting in the following expression: 
\begin{align}\label{eq:dan}
\begin{split}
x^{(j+1)} &= \argmin_x \norm{y - (x * k^{(j)})\downarrow_s}_1\\
k^{(j+1)} &= \argmin_k \norm{y - (x^{(j+1)} * k)\downarrow_s}_1
\end{split}
\end{align}
where $j$ presents the iteration round, $j\in[1,J]$. 
Both components are trained using the sum of the absolute difference, $L_1$ loss, between $k$ and $\hat{k}$, and between $x$ and $\hat{x}$ estimated by the last iteration.

For multi-frame experiments, as shown in blue in Fig.~\ref{fig:overview}, we used the LR feature maps at the last \textit{restorer} iteration before upsampling and adopted EDVR's \textit{PCD Module}, \textit{TSA Module}, and \textit{Restoration Module} for temporal alignment, fusion, and video restoration respectively.
In other words, we merged kernel estimation and blind image restoration techniques with MFSR motion compensation methods and made alterations in order for these modules to utilize temporal kernel consistency. 
Further details of these modules and the architecture can be found in the supplementary material (s.m. Sec.~2 \& s.m. Fig.~1).

\noindent \textbf{Training Data.}
We combined both REDS \cite{REDS} training and validation set and randomly sampled 250 for train and 10 for
test. Following~\cite{KernelGAN}, we generated anisotropic Gaussian kernels with a size of 13$\times$13. 
The lengths of both axes were uniformly sampled in (0.6, 5), and then rotated by a random angle uniformly distributed in [-$\pi$, $\pi$].
For real-world videos, we further added uniform multiplicative noise, up to 25\% of each pixel value of the kernel, to the generated noise-free kernel, and normalized it to sum to one.
Each frame of each HR video was degraded with a randomly generated kernel and then downsampled using bicubic interpolation to form the synthetic LR videos. 
Following previous works~\cite{SRMD, IKC, DAN}, we reshaped the kernels and reduced them through principal component analysis (PCA) before feeding into the network.
We adopted this frame-wise synthesis approach for two reasons:
1) to the best of our knowledge, there is no video dataset with real-world kernels available, and extracting large amount of kernel sequences from video benchmarks for training is costly. 
2) the synthetic training kernels generated as mentioned above can create various degradation in the individual frames, and thus are able to model real-world videos with varying levels of kernel temporal consistency. 

\noindent \textbf{Testing Data.}
We created our testing set with 10 sequences from the REDS testing set (000 and 010-018), denoted as \textit{REDS10}, aiming to mimic the actual degradation of real-world videos that are of varying video dynamicity. 
Concretely, following our experiments in Sec.~\ref{sec:tke}, we first sampled videos from the Something-Something dataset~\cite{something}\footnote{In particular, sequences 13, 16, 21, 35, 37, 49, 52, 55, 63, and 71.}. The sequences from Something-Something dataset
were randomly sampled such that their estimated kernels
had differing temporal kernel consistency. These kernels
were then used to degrade our test set to mimic the degradation characteristics of real-world videos.
We then randomly sampled a sequence from these estimated real-world kernel sequences and used it to downsample each selected video in \textit{REDS10}\footnote{For cases in which the length of the video is longer than the selected kernel sequence, we loop over the same kernel sequence for the remaining frames.}.
As a result, our testing set has the similar degradation characteristics as that of real-world videos, while allow us to perform quantitative evaluations.
The kernel temporal consistency of this test set can be found in the supplementary material (s.m. Fig.~2).
For real-world video evaluations, we used videos from the Something-Something dataset. 
All implementation details can be found in the supplementary material (s.m. Sec.~3).
\begin{figure}[t]
 \centering
 \includegraphics[width=\columnwidth]{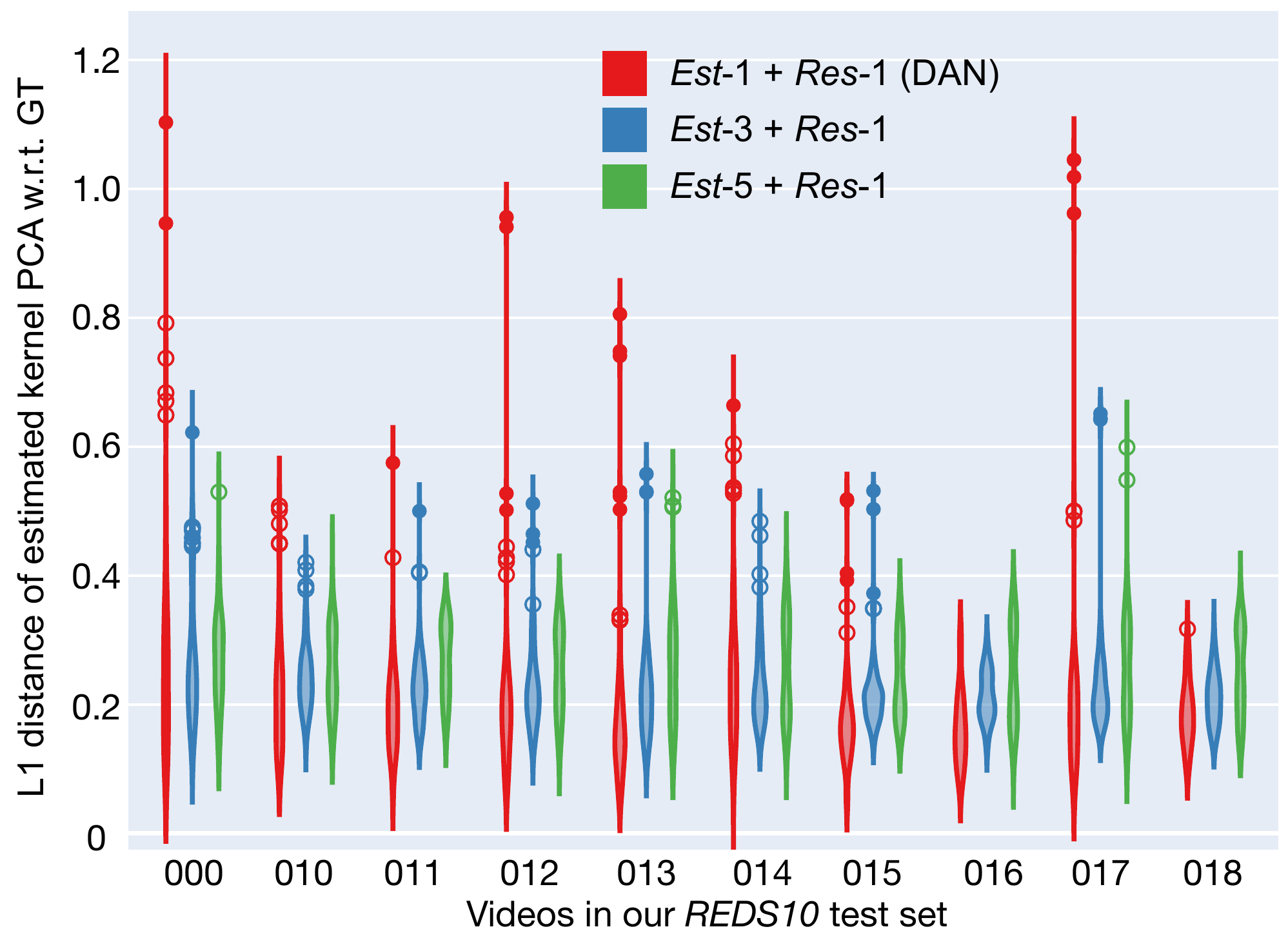}
 \vspace{-5mm}
\caption{Distribution of kernel estimation errors of different \textit{estimators} for each video sequence in our test set. Single-frame \textit{estimator} (\textit{Est}-1+\textit{Res}-1) tends to perform worse than multi-frame estiamtors (\textit{Est}-3/5+\textit{Res}-1) by having larger error variance with many outliers.}
\vspace{-5mm}
 \label{fig:kernel_diffs}
\end{figure}

\subsection{Effectiveness of Temporal Kernel Consistency}\label{sec:tke_effectiveness}
\noindent \textbf{Temporal Kernel Estimation.}
We first studied the effectiveness of taking multiple frames into account for kernel estimation.
In other words, instead of estimating kernels individually for each frame, we leveraged our key insight that the downsampling kernels of frames within a video are temporally consistent to achieve a faster and more accurate kernel estimation for videos.
To this end, we modified the \textit{estimator} to take in multiple LR frames, $\{y_{t+i}\}^{i=N}_{i=-N}$, and generated their corresponding estimated kernels, $\{\hat{k}^{(j)}_{t+i}\}^{i=N}_{i=-N}$.
We then utilized the existing channel attention block in DAN by adopting an early fusion approach, which merges information at the beginning of the block, to exploit the inter-channel relationships not only between basic and conditional inputs, but also among temporal inputs.
Specifically, the features of the HR frames are concatenated with the LR features in every CRB in order to leverage the existing structure of DAN's \textit{estimator} without adding additional channels or layers as shown in the supplementary material (s.m. Fig.~1).

We experimented with different number of input frames on the \textit{estimator}, labelled as \textit{Est-$\alpha$} where \textit{$\alpha$} is the number of frames used for kernel estimation.
Likewise, we labelled $\beta$ as the number of frames used for restoration, \textit{Res-$\beta$}.
For a fair comparison, here we used DAN's \textit{restorer}, $\beta=1$, which is single-frame and therefore not including our adopted EDVR components.
Fig.~\ref{fig:kernel_diffs} shows the distribution of kernel estimation errors of the aforementioned models in terms of the absolute sum of PCA difference between the estimated kernels and their respective ground truth kernels for all frames in each sequence found in \textit{REDS10}.
We observed that independent kernel estimation per-frame can lead to a larger variance and numerous outliers as compared to temporal kernel estimation. 
Notably, temporal kernel estimation results in, on average, more accurate kernels for videos with high dynamicity, \textit{i.e.} low kernel temporal consistency, while performs similarly for videos with high kernel temporal consistency.
The performance increase in kernel estimation, however, did not improve performance significantly in video restoration as shown in Table~\ref{tab:mc_mfr}.
This phenomenon is also observed in recent blind iterative image SR works~\cite{IKC, DAN} and these works reported that this was due to the \textit{restorer}'s robustness to the kernel estimation errors of the \textit{estimator} since they were jointly trained. 
Although having a more accurate kernel estimation did not drastically impact a single-frame video restoration performance, we show that it is essential at improving the performance of a multi-frame restoration approach.

\begin{table}[t!]
\centering
\begin{tabular}{@{}|l|l|@{}}
\hline
Models                               & PSNR/SSIM     \\\hline
\textit{Est}-1 + \textit{Res}-1 (DAN)  & 26.28/0.7118  \\\hline
\textit{Est}-3 + \textit{Res}-1       & 26.30/0.7124 \\
\textit{Est}-5 + \textit{Res}-1       & 26.31/0.7213 \\\hline
\textit{Est}-1 + \textit{Res}-3       & 26.37/0.7170 \\
\textit{Est}-1 + \textit{Res}-5       & 26.54/0.7287\\\hline
\textbf{\textit{Est}-3 + \textit{Res}-3} & \textcolor{blue}{\textbf{26.62/0.7364}}  \\
\textbf{\textit{Est}-5 + \textit{Res}-5} & \textcolor{red}{\textbf{26.76/0.7400}}  \\
\hline
\end{tabular}
\caption{Ablation study on the impact of utilizing temporal kernel estimation on video restoration for both kernel estimation and motion compensation. Red for best performing model and blue for second best. Although estimating more accurate kernels did not significantly improve the performance of a single-image restorer, it is critical for motion compensation and hence benefiting multi-frame restorers.}
\label{tab:mc_mfr}
\vspace{-5mm}
\end{table}


\noindent \textbf{Incorporating Temporal Kernels for MFSR.}
The performance gain of utilizing the temporal information of multiple frames is dependent on the accuracy of its motion estimation; an inaccurate flow can result in misaligned frames after motion compensation and thus artifacts in the restored video~\cite{Tao2017, Xue2018, TDAN, Jo2018}.
As shown in Sec.~\ref{sec:importance_tke}, performing motion compensation under the assumption of a fixed SR kernel directly on real-world videos can result in regular artifacts in the warped frames.
To mitigate this, instead of following the convention of employing motion compensation on the LR frames or features directly, we performed motion compensation on the LR frames \textit{after} considering their corresponding kernels. 
Specifically, we utilized the feature maps at the last \textit{restorer} iteration as shown in Fig.~\ref{fig:overview} which embed both LR frame and the corresponding kernel features from the \textit{estimator}, and then adopted EDVR for temporal alignment, fusion, and restoration as mentioned in Sec.~\ref{sec:setup}.
This approach mitigates the problem of inaccurate motion compensation caused by kernel variation in real-world videos, but the restoration performance may still depend on the accuracy of estimated kernels; errors in kernel estimation would propagate and result in inaccurate motion compensation.

\begin{table}[t!]
\centering
\begin{tabular}{@{}|l|l|l|@{}}
\hline
Proposed for & Models & PSNR/SSIM \\ \hline
MFSR & TDAN~\cite{TDAN} & 25.93/0.6867 \\ 
&EDVR~\cite{EDVR} & 26.21/0.7060 \\ \hline
Blind SISR &IKC~\cite{IKC} & 26.22/0.7021 \\ 
&DAN~\cite{DAN} & 26.28/0.7118 \\ \hline
Blind MFSR & DBVSR~\cite{Pan2020} & 26.11/0.6986 \\ 
&\textbf{\textit{Est}-3 + \textit{Res}-3 (Ours)} & \textcolor{blue}{\textbf{26.62}/\textbf{0.7364}} \\ 
&\textbf{\textit{Est}-5 + \textit{Res}-5 (Ours)} & \textcolor{red}{\textbf{26.76}/\textbf{0.7400}} \\ \hline
\end{tabular}
\caption{We compare our model with state-of-the-art models from MFSR, which assume a fixed bicubic degradation, and blind single-image SR methods, which restore each frame independently.}
\vspace{-7mm}
\label{tab:compare}
\end{table}

To verify this, we first ran our multi-frame \textit{restorer}, $\beta=\{3,5\}$, with a single-frame \textit{estimator}, $\alpha=1$ and compared it with running the multi-frame \textit{restorer} together with the multi-frame \textit{estimator}.
The results are shown in Table~\ref{tab:mc_mfr}.
As expected, having a multi-frame \textit{restorer} resulted in an improvement in video restoration similar to that of previous works~\cite{Liao2015, Kappeler2016, Liu2017, Makansi2017, Tao2017}.
However, these per-frame \textit{estimator} MFSR models did not perform as well as their temporal estimator counterparts.
In particular, although our per-frame \textit{estimator} MFSR model utilized information from $5$ frames (\textit{Est}-1 + \textit{Res}-5) to restore each frame, it did not outperform our temporal \textit{estimator} MFSR model that only exploited information from $3$ frames (\textit{Est}-3 + \textit{Res}-3).
Hence, we can conclude that the kernel mismatch errors incurred during kernel estimation propagated through the implicit motion compensation module of EDVR, affecting temporal alignment, fusion, and thus restoration.
In other words, more accurate estimated kernels through the temporal kernel \textit{estimator} enable the multi-frame \textit{restorer} to leverage temporal frame information better.
Therefore, the interplay between accurate kernel estimation and motion compensation is the key to utilize temporal kernel consistency for video restoration. 

\begin{figure*}[t!]
 \centering
 \includegraphics[width=\columnwidth]{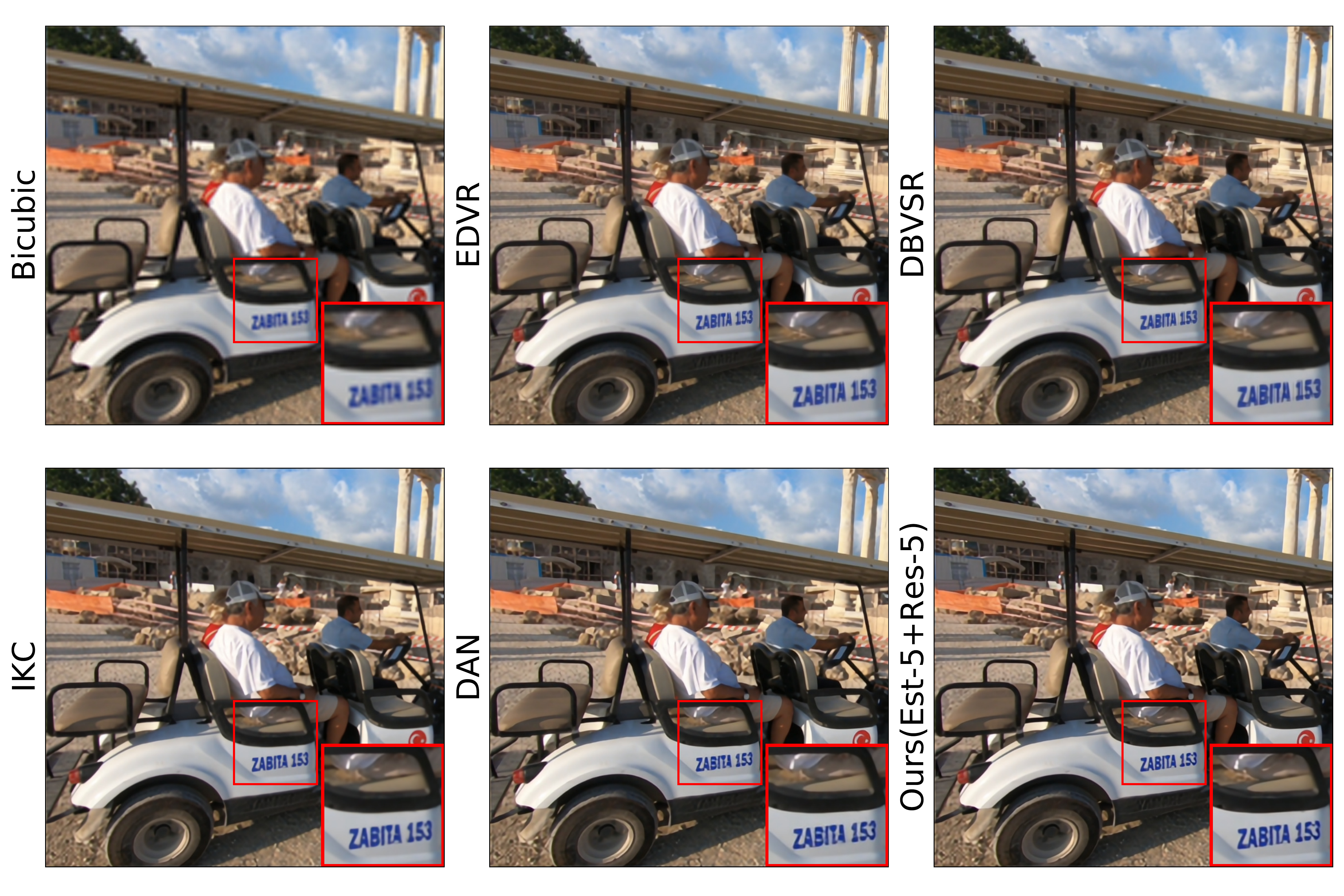}
 \includegraphics[width=\columnwidth]{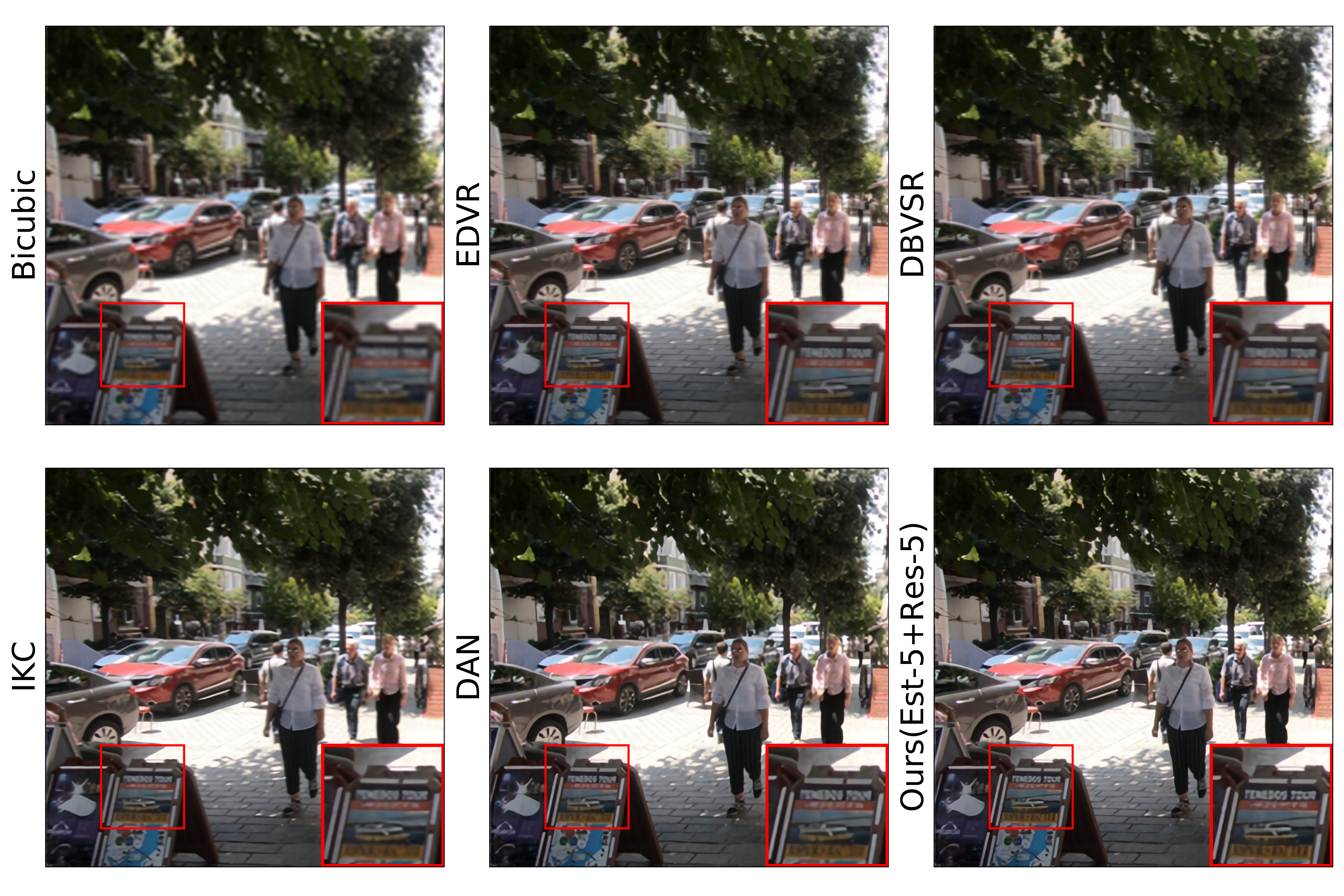}\\
 \includegraphics[width=\columnwidth]{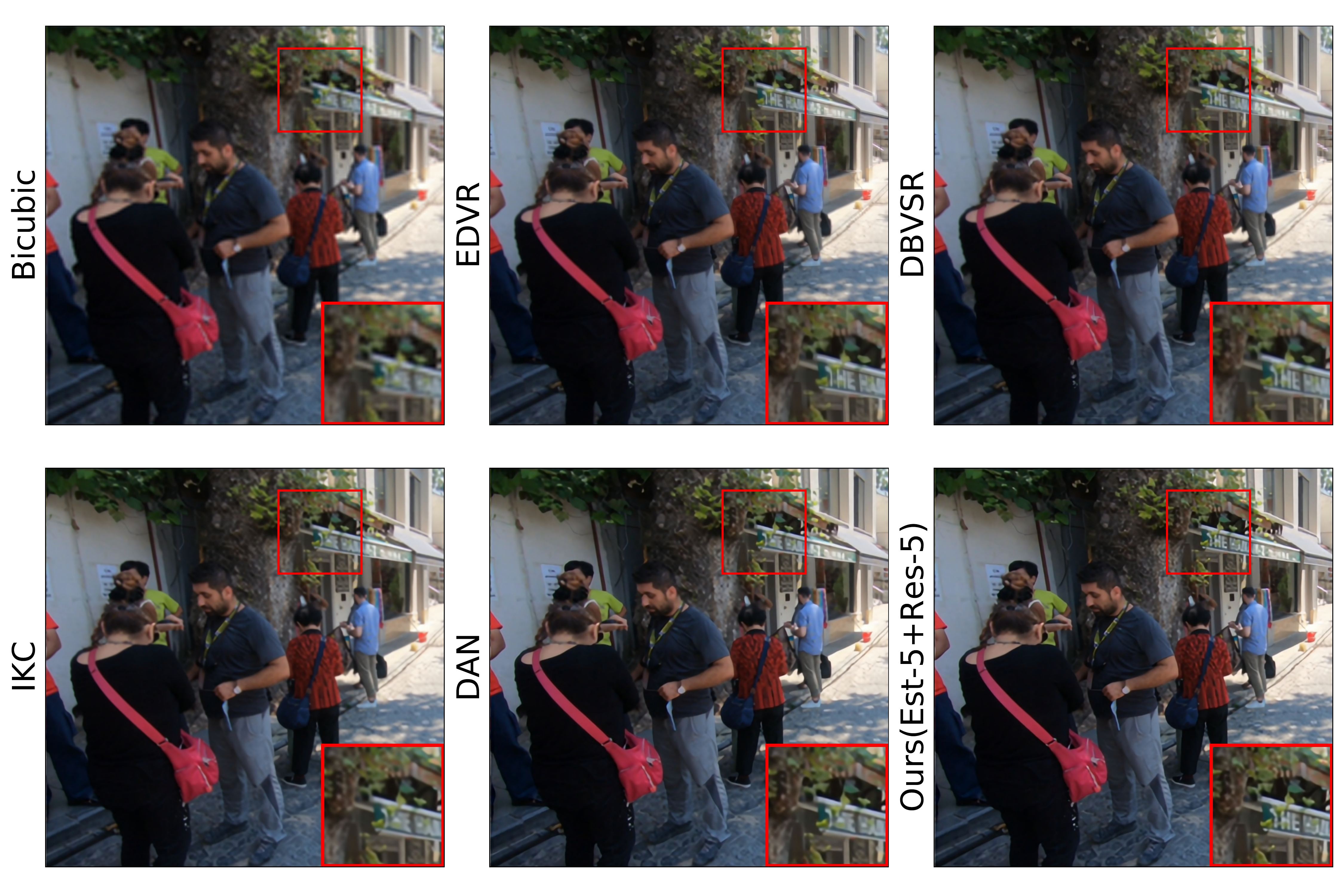} 
 \includegraphics[width=\columnwidth]{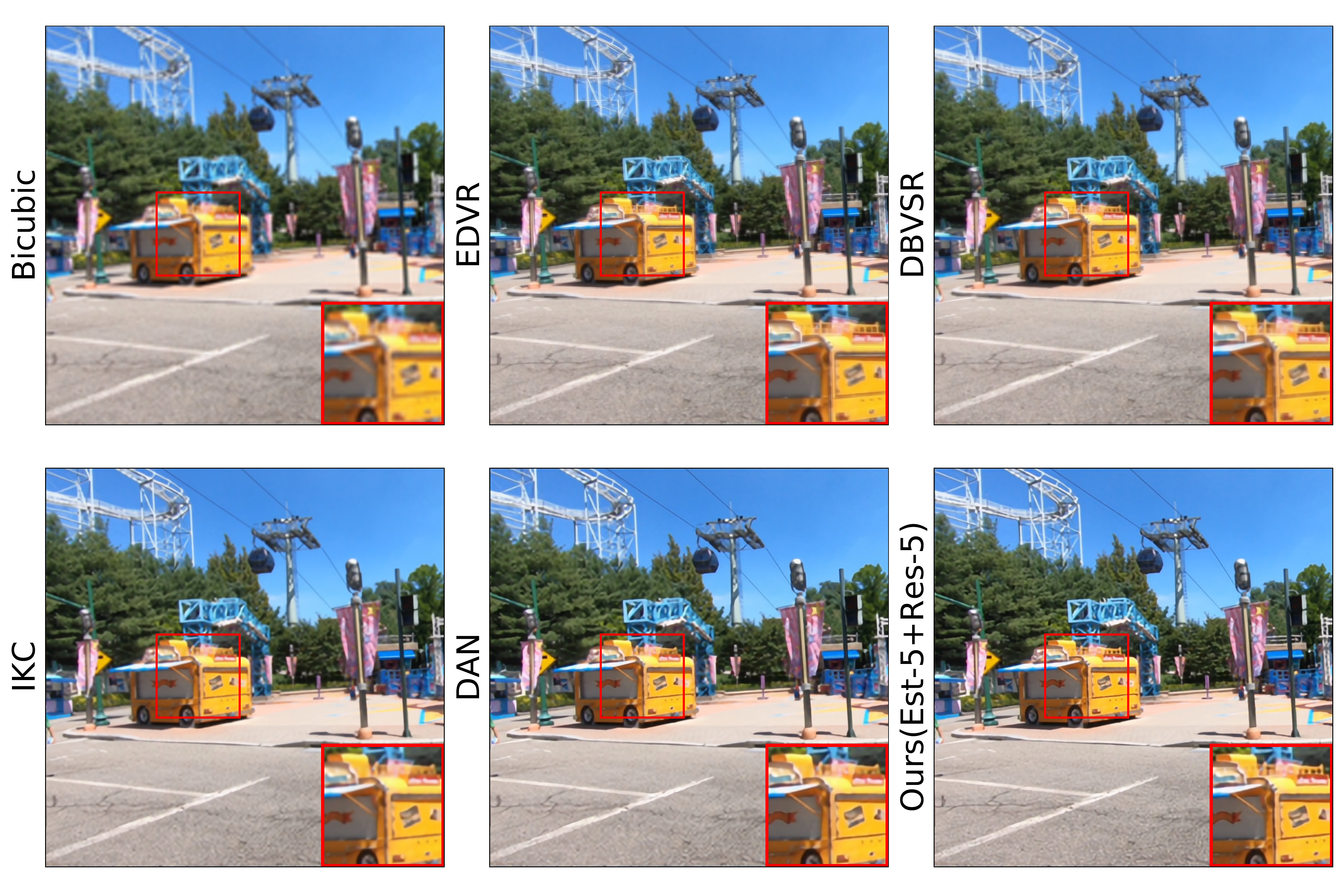}
 \caption{Qualitative comparison among existing models, along with bicubic upscaling, on our benchmark test sequences. Zoom in for best results.}
 \vspace{-4mm}
 \label{fig:benchmark}
\end{figure*}

\begin{figure*}[ht!]
 \centering
 \includegraphics[width=\columnwidth]{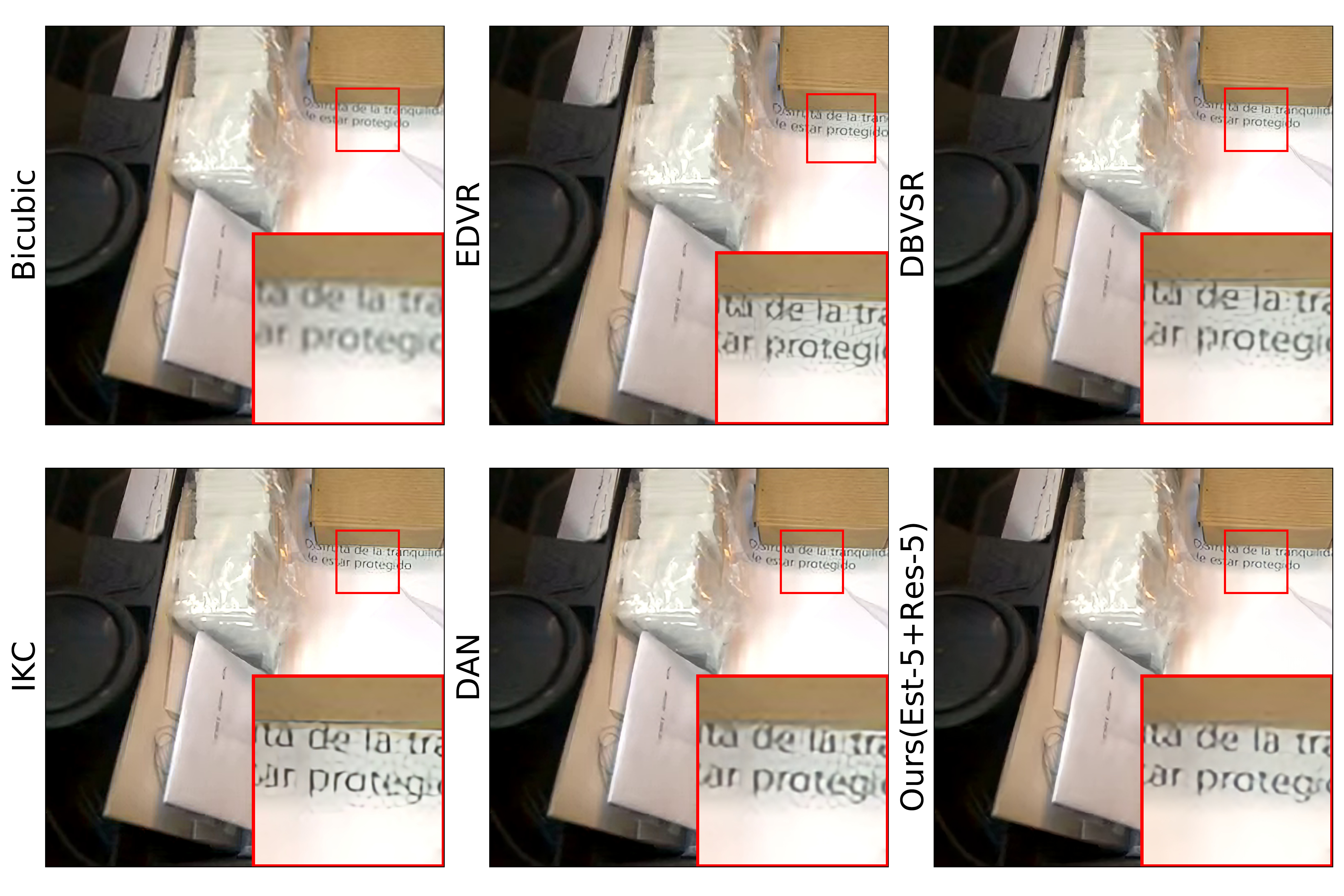}
 \includegraphics[width=\columnwidth]{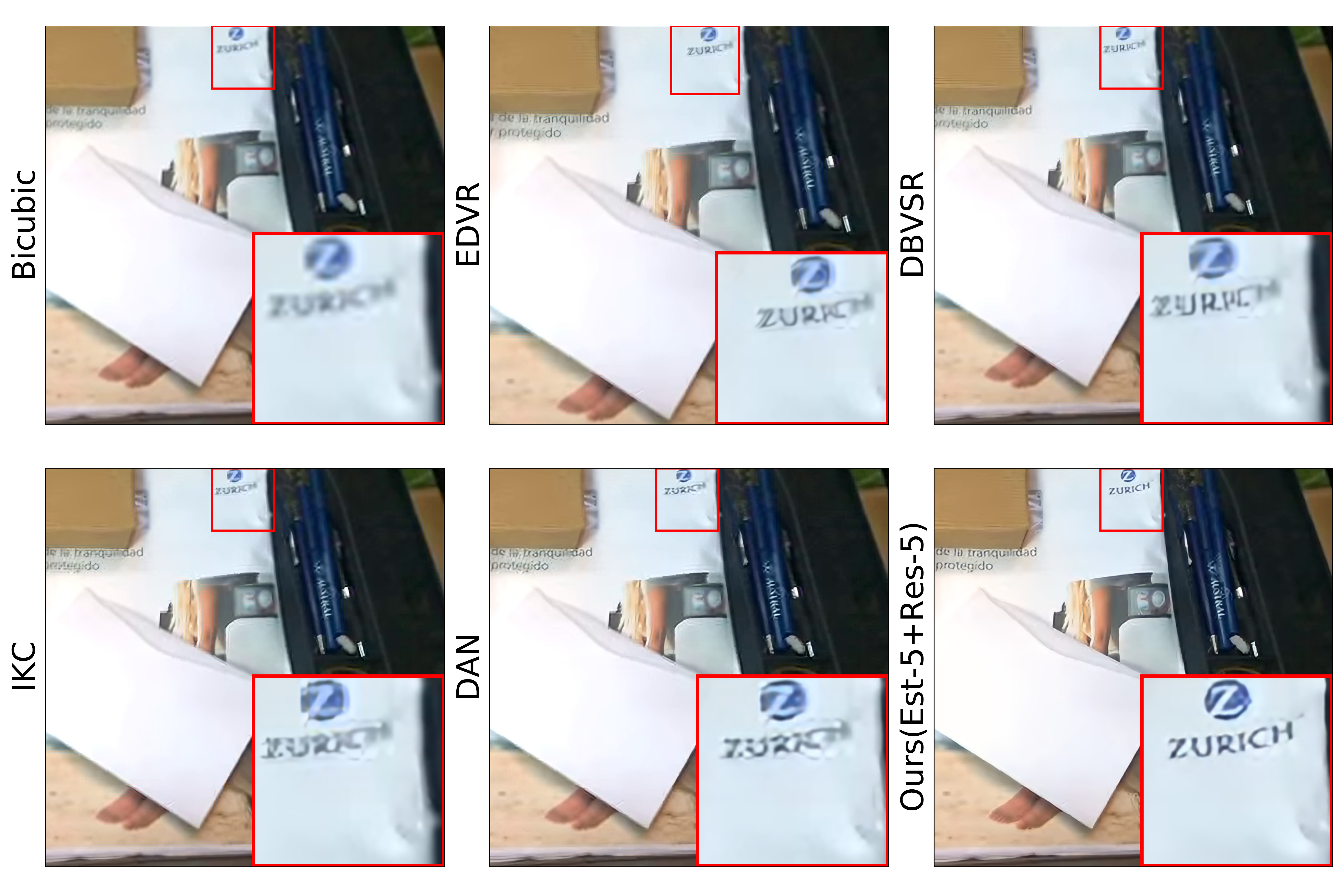}
 \caption{Real-world qualitative comparisons among existing models, along with bicubic upscaling. Zoom in for best results. Note that there is no ground-truth available.}
 \vspace{-2mm}
 \label{fig:realworld}
\end{figure*}

\noindent \textbf{Comparisons with Previous Works.}
We compared our approach, with existing works on both our test set \textit{REDS10} and real-world videos taken from the Something-Something dataset.
Specifically, we considered state-of-the-art MFSR methods, TDAN~\cite{TDAN} and EDVR~\cite{EDVR}, blind image-based SR methods, IKC~\cite{IKC} and DAN~\cite{DAN}, and a recently proposed blind MSFR approach, DBVSR~\cite{Pan2020}.

From the quantitative and qualitative comparisons based on \textit{REDS10} (Table.~\ref{tab:compare} \& Fig.~\ref{fig:benchmark}) and real-world qualitative examples (Fig.~\ref{fig:realworld}), we observe that existing MFSR approaches are lacking due to kernel mismatch, affecting both motion compensation and video restoration as shown in Sec.~\ref{sec:importance_tke}.
Both TDAN and EDVR, in particular, were trained using the fixed bicubic degradation assumption and DBVSR assumed a fixed temporally uniform kernel.
Blind SISR approaches, on the other hand, restore each frame independently and hence perform slightly better than existing MFSR approaches.
Our approach, which exploits kernel temporal consistency for accurate kernel estimation and mitigates the effects of kernel mismatch on motion compensation, leads to a dominant solution for real-world video restoration.
We provided additional examples in the supplementary material (s.m. Fig.~5 \& Fig.~6). 

\section{Conclusion}

\vspace{-3mm}
In this paper, we presented the temporal kernel changes in videos and showed that they varied in their consistency depending on the video's dynamicity.
Through our experiments, we highlighted the importance of estimating kernels per-frame to tackle the effects of temporal kernel mismatch in previous works.
We then showed how temporal kernel consistency can be generally incorporated into existing works through the interaction between both kernel estimation and motion compensation in order to leverage both temporal kernel and frame information for blind video SR.
We hope to influence future blind video SR model design by emphasizing the potential of leveraging kernel temporal consistency in restoring videos.


{\small
\bibliographystyle{ieee_fullname}
\bibliography{99_main}

\begin{thebibliography}{10}\itemsep=-1pt

\bibitem{CARN}
Namhyuk Ahn, Byungkon Kang, and Kyung-Ah Sohn.
\newblock {Fast, Accurate, and Lightweight Super-Resolution with Cascading
  Residual Network}.
\newblock In {\em ECCV}, 2018.

\bibitem{Baker2007}
S. Baker, D. Scharstein, J. Lewis, S. Roth, Michael~J. Black, and R. Szeliski.
\newblock A database and evaluation methodology for optical flow.
\newblock {\em International Journal of Computer Vision}, 2007.

\bibitem{KernelGAN}
Sefi Bell-Kligler, Assaf Shocher, and Michal Irani.
\newblock Blind super-resolution kernel estimation using an internal-gan.
\newblock In {\em Advances in Neural Information Processing Systems}. 2019.

\bibitem{Cai2019}
Jianrui Cai, Hui Zeng, Hongwei Yong, Zisheng Cao, and Lei Zhang.
\newblock Toward real-world single image super-resolution: A new benchmark and
  a new model.
\newblock {\em IEEE/CVF International Conference on Computer Vision (ICCV)},
  2019.

\bibitem{Dai2017}
Jifeng Dai, Haozhi Qi, Y. Xiong, Y. Li, Guodong Zhang, H. Hu, and Y. Wei.
\newblock Deformable convolutional networks.
\newblock {\em IEEE International Conference on Computer Vision (ICCV)}, 2017.

\bibitem{SRCNN}
Chao Dong, Chen~Change Loy, Kaiming He, and Xiaoou Tang.
\newblock Image super-resolution using deep convolutional networks.
\newblock {\em IEEE Transactions on Pattern Analysis and Machine Intelligence},
  2016.

\bibitem{FSRCNN}
Chao Dong, Chen~Change Loy, and Xiaoou Tang.
\newblock {Accelerating the Super-Resolution Convolutional Neural Network}.
\newblock In {\em ECCV}, 2016.

\bibitem{Efrat2013}
N. Efrat, Daniel Glasner, Alexander Apartsin, B. Nadler, and A. Levin.
\newblock Accurate blur models vs. image priors in single image
  super-resolution.
\newblock {\em IEEE International Conference on Computer Vision (ICCV)}, 2013.

\bibitem{Farsiu2004}
Sina Farsiu, M.~D. Robinson, Michael Elad, and P. Milanfar.
\newblock Fast and robust multiframe super resolution.
\newblock {\em IEEE Transactions on Image Processing}, 2004.

\bibitem{Goodfellow2014}
Ian~J. Goodfellow, Jean Pouget-Abadie, M. Mirza, Bing Xu, David Warde-Farley,
  Sherjil Ozair, Aaron~C. Courville, and Yoshua Bengio.
\newblock Generative adversarial nets.
\newblock In {\em Advances in Neural Information Processing Systems}, 2014.

\bibitem{something}
Raghav Goyal, Samira~Ebrahimi Kahou, Vincent Michalski, Joanna Materzynska,
  Susanne Westphal, Heuna Kim, Valentin Haenel, Ingo Fruend, Peter Yianilos,
  Moritz Mueller-Freitag, et~al.
\newblock {The Something Something Video Database for Learning and Evaluating
  Visual Common Sense.}
\newblock In {\em ICCV}, 2017.

\bibitem{IKC}
Jinjin Gu, Hannan Lu, W. Zuo, and C. Dong.
\newblock Blind super-resolution with iterative kernel correction.
\newblock {\em IEEE/CVF Conference on Computer Vision and Pattern Recognition
  (CVPR)}, 2019.

\bibitem{DRN}
Yong Guo, Jian Chen, J. Wang, Q. Chen, Jiezhang Cao, Zeshuai Deng, Yanwu Xu,
  and Mingkui Tan.
\newblock Closed-loop matters: Dual regression networks for single image
  super-resolution.
\newblock {\em IEEE/CVF Conference on Computer Vision and Pattern Recognition
  (CVPR)}, 2020.

\bibitem{FlowNet2}
Eddy Ilg, N. Mayer, Tonmoy Saikia, Margret Keuper, A. Dosovitskiy, and T. Brox.
\newblock Flownet 2.0: Evolution of optical flow estimation with deep networks.
\newblock {\em IEEE Conference on Computer Vision and Pattern Recognition
  (CVPR)}, 2017.

\bibitem{Jaderberg2015}
Max Jaderberg, K. Simonyan, Andrew Zisserman, and K. Kavukcuoglu.
\newblock Spatial transformer networks.
\newblock In {\em Advances in Neural Information Processing Systems}, 2015.

\bibitem{Jo2018}
Younghyun Jo, S. Oh, Jaeyeon Kang, and S. Kim.
\newblock Deep video super-resolution network using dynamic upsampling filters
  without explicit motion compensation.
\newblock {\em IEEE/CVF Conference on Computer Vision and Pattern Recognition},
  2018.

\bibitem{Kappeler2016}
Armin Kappeler, Seunghwan Yoo, Qiqin Dai, and A. Katsaggelos.
\newblock Video super-resolution with convolutional neural networks.
\newblock {\em IEEE Transactions on Computational Imaging}, 2016.

\bibitem{kingma2014adam}
Diederik~P Kingma and Jimmy Ba.
\newblock Adam: A method for stochastic optimization.
\newblock {\em arXiv preprint arXiv:1412.6980}, 2014.

\bibitem{TPSR}
Royson Lee, L. Dudziak, M. Abdelfattah, Stylianos~I. Venieris, H. Kim, Hongkai
  Wen, and N. Lane.
\newblock Journey towards tiny perceptual super-resolution.
\newblock In {\em European Conference on Computer Vision (ECCV)}, 2020.

\bibitem{Liao2015}
Renjie Liao, X. Tao, R. Li, Z. Ma, and J. Jia.
\newblock Video super-resolution via deep draft-ensemble learning.
\newblock {\em IEEE International Conference on Computer Vision (ICCV)}, 2015.

\bibitem{Liu2014}
Ce Liu and Deqing Sun.
\newblock On bayesian adaptive video super resolution.
\newblock {\em IEEE Transactions on Pattern Analysis and Machine Intelligence},
  2014.

\bibitem{Liu2017}
Ding Liu, Zhaowen Wang, Yuchen Fan, X. Liu, Zhangyang Wang, S. Chang, and T.
  Huang.
\newblock Robust video super-resolution with learned temporal dynamics.
\newblock {\em IEEE International Conference on Computer Vision (ICCV)}, 2017.

\bibitem{DAN}
Zhengxiong Luo, Y. Huang, Shang Li, Liang Wang, and Tieniu Tan.
\newblock Unfolding the alternating optimization for blind super resolution.
\newblock In {\em Advances in Neural Information Processing Systems}. 2020.

\bibitem{Ma2015}
Z. Ma, Renjie Liao, X. Tao, L. Xu, J. Jia, and Enhua Wu.
\newblock Handling motion blur in multi-frame super-resolution.
\newblock {\em IEEE Conference on Computer Vision and Pattern Recognition
  (CVPR)}, 2015.

\bibitem{Makansi2017}
Osama Makansi, Eddy Ilg, and T. Brox.
\newblock End-to-end learning of video super-resolution with motion
  compensation.
\newblock In {\em German Conference on Pattern Recognition (GCPR)}, 2017.

\bibitem{Michaeli2013}
T. Michaeli and M. Irani.
\newblock Nonparametric blind super-resolution.
\newblock {\em IEEE International Conference on Computer Vision (ICCV)}, 2013.

\bibitem{REDS}
Seungjun Nah, Sungyong Baik, Seokil Hong, Gyeongsik Moon, Sanghyun Son, Radu
  Timofte, and Kyoung~Mu Lee.
\newblock Ntire 2019 challenge on video deblurring and super-resolution:
  Dataset and study.
\newblock In {\em The IEEE Conference on Computer Vision and Pattern
  Recognition (CVPR) Workshops}, 2019.

\bibitem{SPynet}
A. Ranjan and Michael~J. Black.
\newblock Optical flow estimation using a spatial pyramid network.
\newblock {\em 2017 IEEE Conference on Computer Vision and Pattern Recognition
  (CVPR)}, 2017.

\bibitem{FRVSR}
Mehdi S.~M. Sajjadi, Raviteja Vemulapalli, and M. Brown.
\newblock Frame-recurrent video super-resolution.
\newblock {\em 2018 IEEE/CVF Conference on Computer Vision and Pattern
  Recognition}, 2018.

\bibitem{Pan2020}
Jin shan Pan, Songsheng Cheng, Jiawei Zhang, and J. Tang.
\newblock Deep blind video super-resolution.
\newblock {\em ArXiv}, 2020.

\bibitem{ESPCN}
W. Shi, J. Caballero, Ferenc Husz{\'a}r, J. Totz, A. Aitken, R. Bishop, D.
  Rueckert, and Zehan Wang.
\newblock Real-time single image and video super-resolution using an efficient
  sub-pixel convolutional neural network.
\newblock {\em IEEE Conference on Computer Vision and Pattern Recognition
  (CVPR)}, 2016.

\bibitem{ZSSR}
Assaf Shocher, N. Cohen, and M. Irani.
\newblock "zero-shot" super-resolution using deep internal learning.
\newblock In {\em IEEE/CVF Conference on Computer Vision and Pattern
  Recognition (CVPR)}, 2018.

\bibitem{ESRN}
Dehua Song, Chang Xu, Xu Jia, Yiyi Chen, Chunjing Xu, and Yunhe Wang.
\newblock Efficient residual dense block search for image super-resolution.
\newblock In {\em Association for the Advancement of Artificial Intelligence
  (AAAI)}, 2020.

\bibitem{PWCNet}
Deqing Sun, X. Yang, Ming-Yu Liu, and J. Kautz.
\newblock Pwc-net: Cnns for optical flow using pyramid, warping, and cost
  volume.
\newblock {\em IEEE/CVF Conference on Computer Vision and Pattern Recognition
  (CVPR)}, 2018.

\bibitem{Tao2017}
X. Tao, H. Gao, Renjie Liao, J. Wang, and J. Jia.
\newblock Detail-revealing deep video super-resolution.
\newblock {\em IEEE International Conference on Computer Vision (ICCV)}, 2017.

\bibitem{TDAN}
Yapeng Tian, Yulun Zhang, Yun Fu, and Chenliang Xu.
\newblock Tdan: Temporally-deformable alignment network for video
  super-resolution.
\newblock {\em IEEE/CVF Conference on Computer Vision and Pattern Recognition
  (CVPR)}, 2020.

\bibitem{EDVR}
Xintao Wang, Kelvin C.~K. Chan, K. Yu, C. Dong, and Chen~Change Loy.
\newblock Edvr: Video restoration with enhanced deformable convolutional
  networks.
\newblock {\em IEEE/CVF Conference on Computer Vision and Pattern Recognition
  Workshops (CVPRW)}, 2019.

\bibitem{SSIM}
Zhou Wang, Alan~C. Bovik, Hamid~R. Sheikh, and Eero~P. Simoncelli.
\newblock {Image quality assessment: from error visibility to structural
  similarity}.
\newblock {\em IEEE Transactions on Image Processing}, 2004.

\bibitem{Xiang2020}
X. Xiang, Yapeng Tian, Yulun Zhang, Y. Fu, J. Allebach, and Chenliang Xu.
\newblock Zooming slow-mo: Fast and accurate one-stage space-time video
  super-resolution.
\newblock {\em IEEE/CVF Conference on Computer Vision and Pattern Recognition
  (CVPR)}, 2020.

\bibitem{Xue2018}
Tianfan Xue, B. Chen, Jiajun Wu, D. Wei, and W. Freeman.
\newblock Video enhancement with task-oriented flow.
\newblock {\em International Journal of Computer Vision}, 2018.

\bibitem{USRNet}
K. Zhang, L. Gool, and R. Timofte.
\newblock Deep unfolding network for image super-resolution.
\newblock {\em IEEE/CVF Conference on Computer Vision and Pattern Recognition
  (CVPR)}, 2020.

\bibitem{SRMD}
Kai Zhang, W. Zuo, and Lei Zhang.
\newblock Learning a single convolutional super-resolution network for multiple
  degradations.
\newblock In {\em IEEE/CVF Conference on Computer Vision and Pattern
  Recognition (CVPR)}, 2018.

\bibitem{RCAN}
Yulun Zhang, Kunpeng Li, Kai Li, Lichen Wang, Bineng Zhong, and Yun Fu.
\newblock {Image Super-Resolution Using Very Deep Residual Channel Attention
  Networks}.
\newblock In {\em European Conference on Computer Vision (ECCV)}, 2018.

\bibitem{KMSR}
Ruofan Zhou and S. S{\"u}sstrunk.
\newblock Kernel modeling super-resolution on real low-resolution images.
\newblock {\em IEEE/CVF International Conference on Computer Vision (ICCV)},
  2019.

\bibitem{Zhu2019}
X. Zhu, H. Hu, Stephen Lin, and Jifeng Dai.
\newblock Deformable convnets v2: More deformable, better results.
\newblock {\em IEEE/CVF Conference on Computer Vision and Pattern Recognition
  (CVPR)}, 2019.

\end{thebibliography}
}

\newpage
\appendix






\renewcommand{\thefigure}{A\arabic{figure}}
\setcounter{figure}{0}
 
\begin{figure*}[t!]
 \centering
 \includegraphics[width=0.95\textwidth]{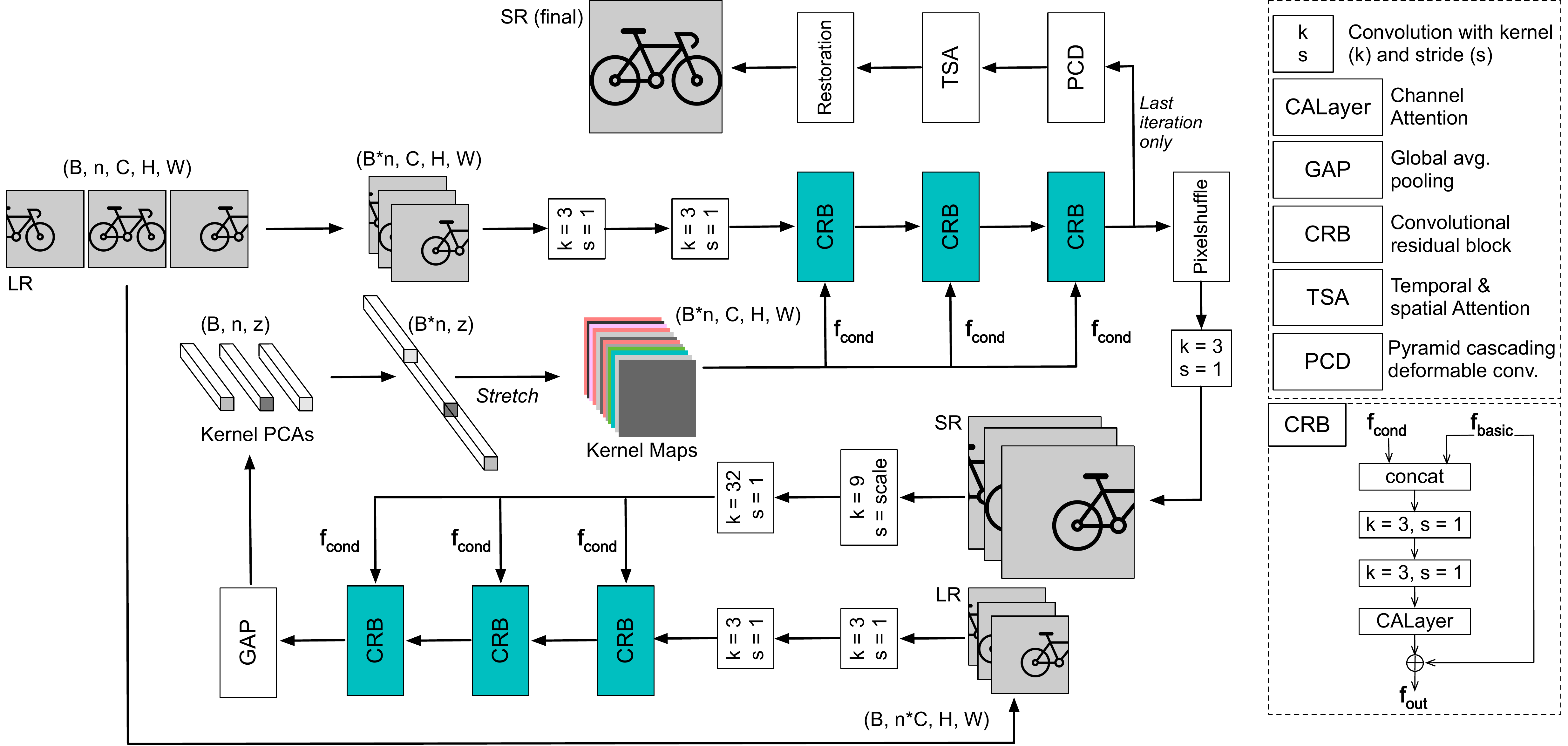}
 \caption{Detailed architecture of our model experimented in Sec.~5 of the main paper.}
 \label{fig:arch}
\end{figure*}

\begin{figure}[t]
 \centering
 \includegraphics[width=\columnwidth]{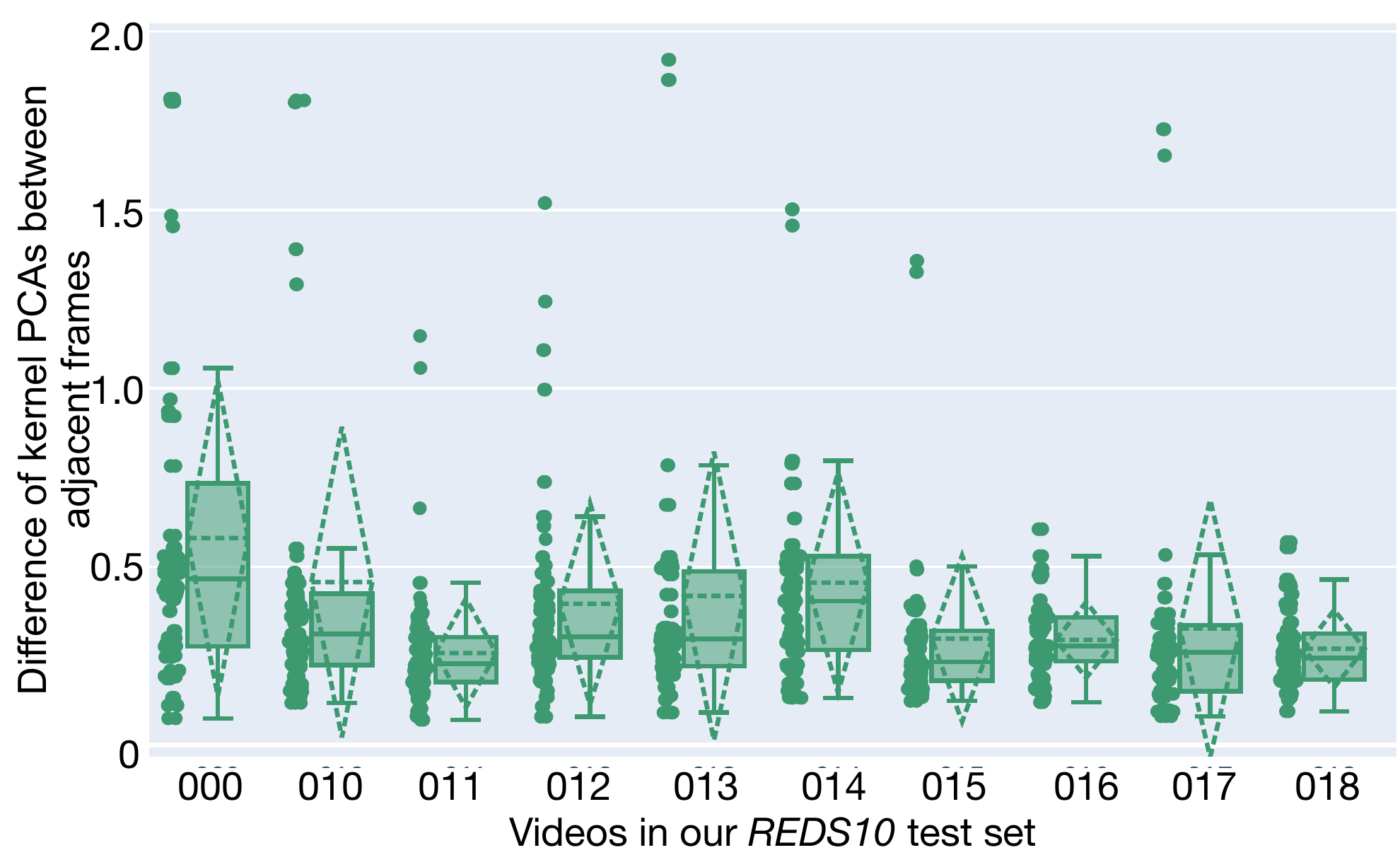}
 \caption{Temporal kernel consistency of videos in our \textit{REDS10} benchmark, measured by kernel PCA changes for adjacent frames in the videos. Kernel changes are represented by solid dots while boxplots show distributions.}
 \label{sup_fig:bench_ktc}
\end{figure}

\section{Implementation Details of KernelGAN \& ZSSR. }

We used the default settings and hyperparameters provided by KernelGAN~\cite{KernelGAN} and ZSSR~\cite{ZSSR}.
For KernelGAN, the estimated downscaling kernel size is set to $13\times13$ and the input image is cropped to $64\times64$ before kernel extraction.
The kernel is extracted after 3000 iterations using the Adam~\cite{kingma2014adam} optimizer with learning rate set to $0.0002$,
$\beta_1$ set to $0.5$ and $\beta_2$ set to $0.999$.
For ZSSR, the input LR image and the provided estimated kernel is used to generate a downsampled variant of the LR image. The resulting image pair is then used to train the model using the Adam optimizer starting with learning rate set to $0.001$ with $\beta_1 = 0.9$ and $\beta_2 = 0.999$.
For more details, please refer to the provided repository\footnote{https://github.com/sefibk/KernelGAN}.

\section{Architecture of DAN \& EDVR}

A more detail architecture of our model experimented in Sec.~5 of the main paper is shown in Fig.~\ref{fig:arch}.
Notably, the features of the HR frames are concatenated with the LR features in each CRB block of DAN~\cite{DAN} and we utilized the existing channel attention layer (CALayer) for temporal kernel estimation.
During the last iteration, the LR features, which were conditioned on the input frames and their estimated kernel, were fed into the temporal blocks of EDVR~\cite{EDVR} for temporal alignment, fusion, and restoration.
In particular, the PCD module follows a pyramid cascading structure, which concatenates features of differing spatial sizes and uses deformable convolution at each respective pyramid level to the aligned features.
The TSA module then fused these aligned features together through both temporal and spatial attention.
Specifically, temporal attention maps are computed based on the aligned features and applied to these features through the dot product before concatenating and fusing them using a convolution layer.
After which, the fused features are then used to compute the spatial attention maps which are then applied to these features.
For more details, please refer to EDVR~\cite{EDVR}.

\section{Implementation Details of DAN \& EDVR. }

For training, we used scaling factor $\times4$, input patch size of 100$\times$100, and set $N=1$, i.e. considering sequences of 3. 
We set $N=2$ to highlight the kernel mismatch on motion compensation as shown in Fig.~\ref{sup_fig:vid_flow}.
The batch size was set to 4, and all models were trained for 300 epochs, using the Adam optimizer~\cite{kingma2014adam} ($\beta_1 = 0.9$ and $\beta_2 = 0.999$).
The initial learning rate was set to $1\times 10^{-4}$, and decayed with a factor of 0.5 at every 200 epochs. 
Following DAN~\cite{DAN}, we ran for 4 iterations ($J=4$) and used L1 loss for both kernel estimation and video restoration across all our models in every iteration. 
When multiple frames were utilized for temporal alignment, we applied a scaling factor of $1/2N$ to weight the loss from supporting frames. 
Following previous works~\cite{EDVR, RCAN}, PSNR and SSIM~\cite{SSIM} were computed after converting each frame from RGB to Y channel and trimming the edges by the scale factor. 
All experiments were run on NVIDIA 1080Ti and 2080Ti GPUs.
Temporal kernel consistency of our test benchmark, \textit{REDS10}, is shown in Fig.~\ref{sup_fig:bench_ktc}. 
Similar to Fig.~1 in the main paper, we quantified kernel temporal change by measuring the sum of absolute difference between consecutive kernel PCAs.
In particular, video sequences such as 016 and 018 have high temporal kernel consistency and sequences such as 000 and 014 have low temporal kernel consistency.

\section{Additional Results}

\begin{figure*}[!ht]
 \centering
 \includegraphics[width=\columnwidth]{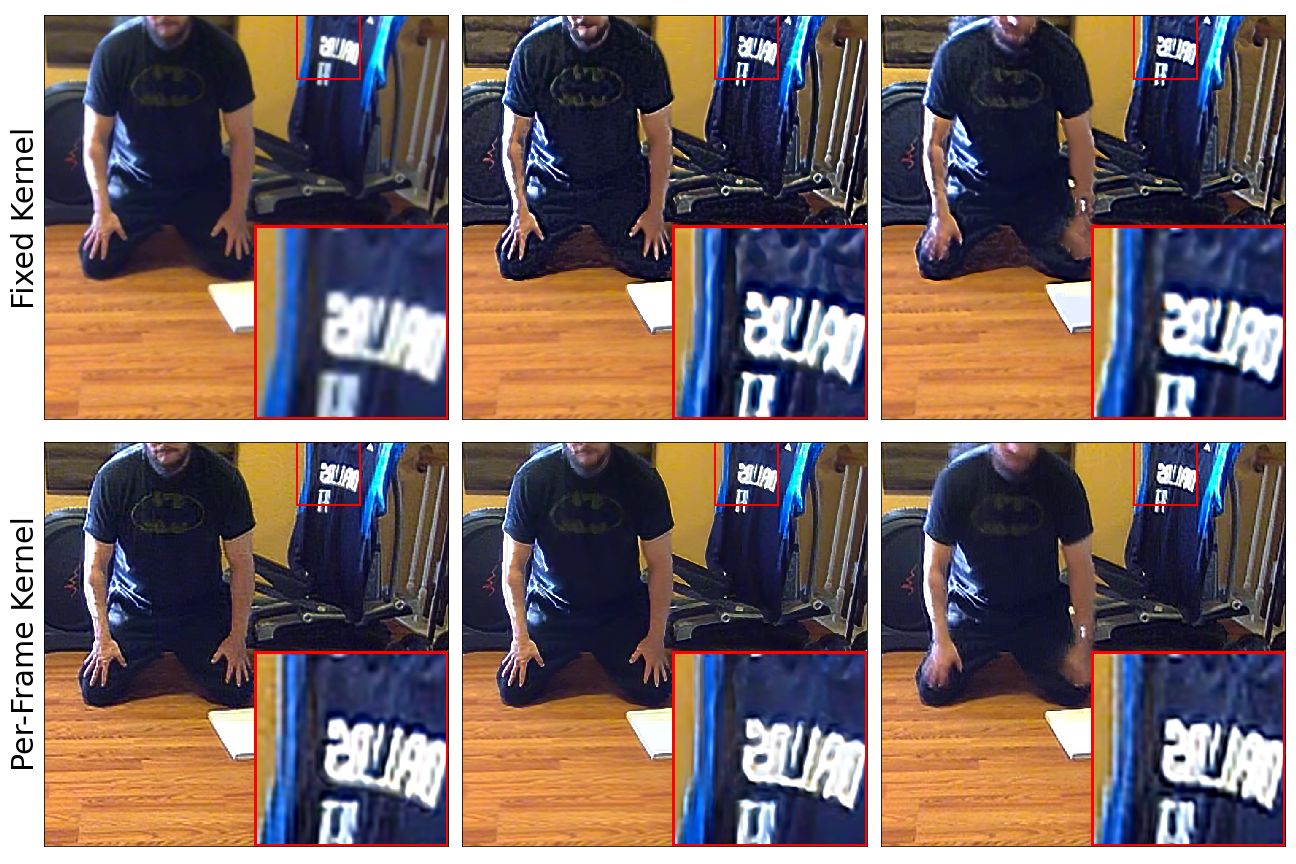}
 \includegraphics[width=\columnwidth ]{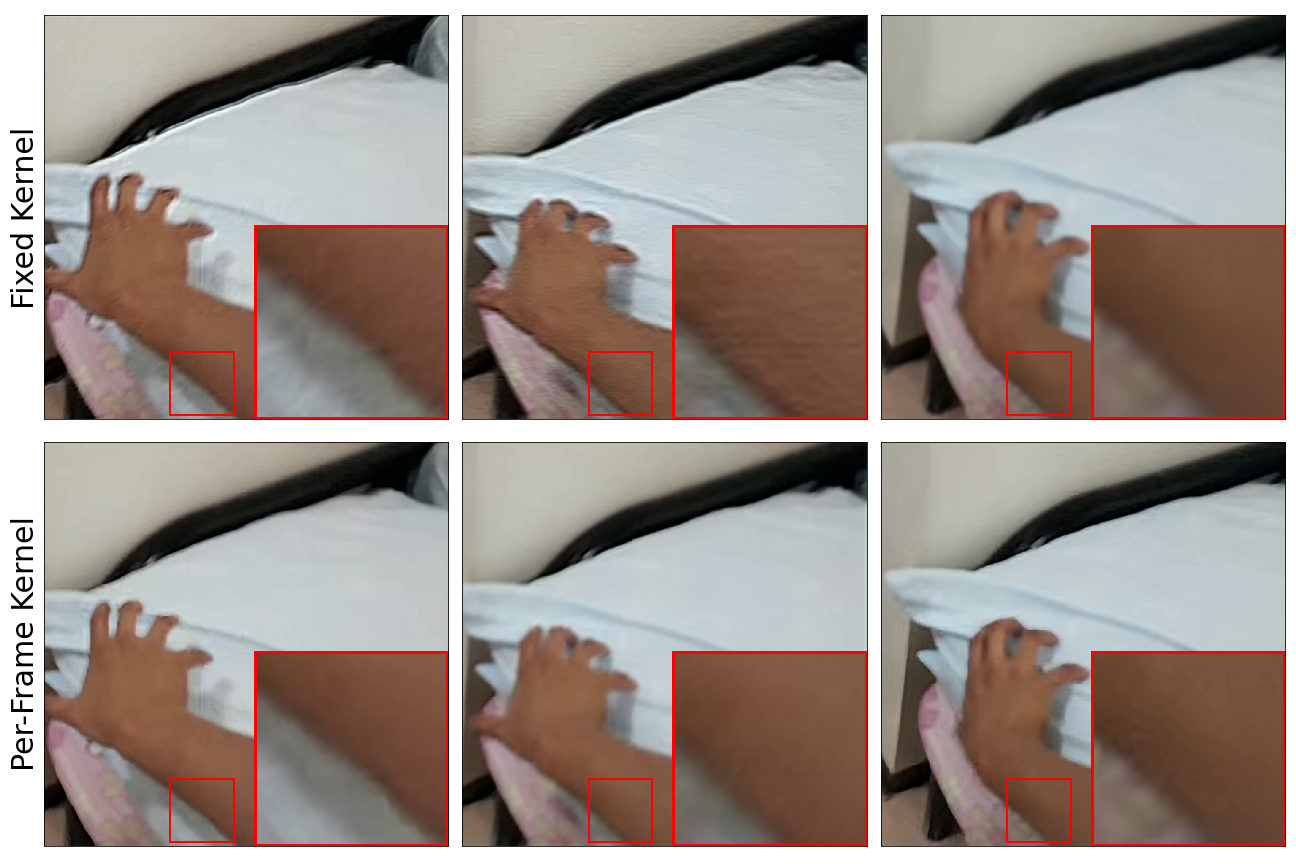}\\
 \includegraphics[width=\columnwidth]{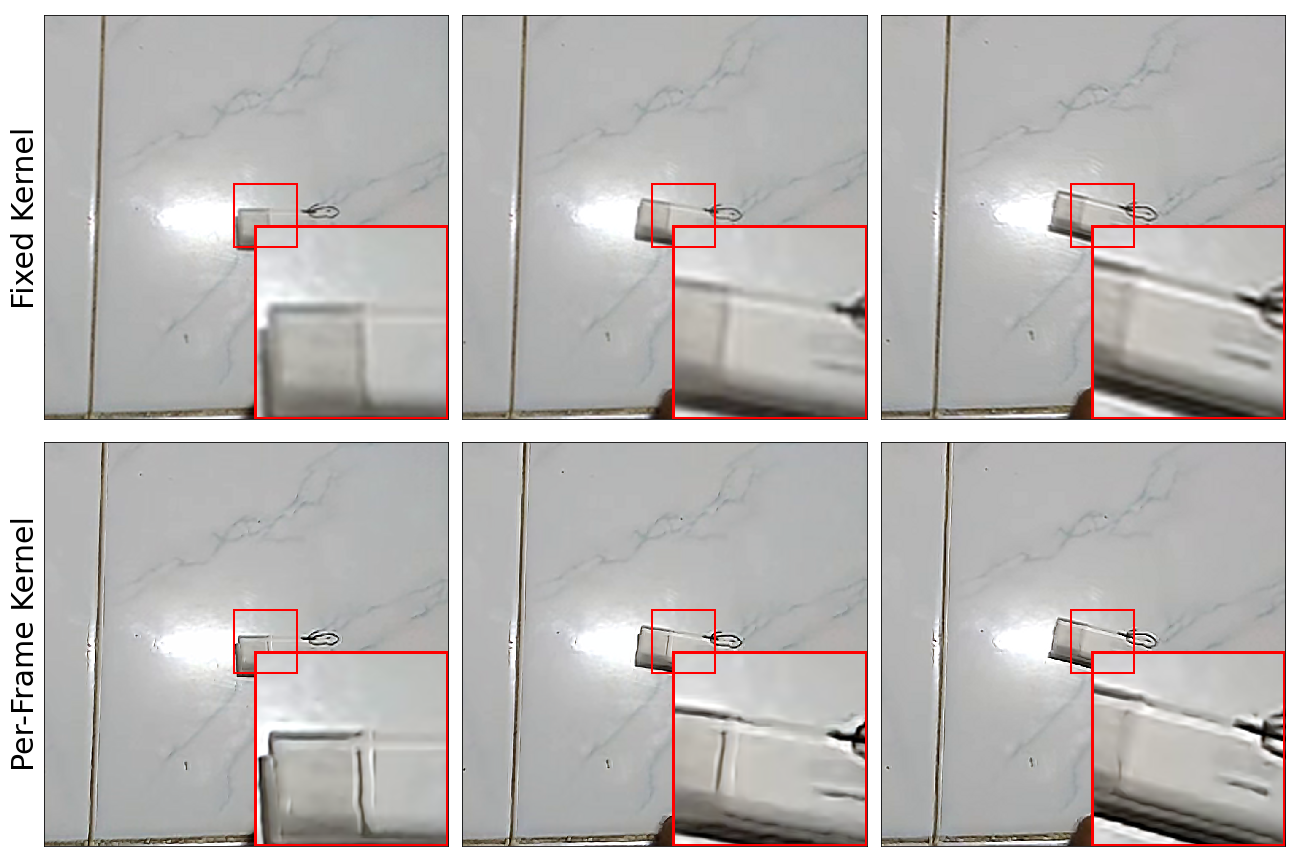}
  \includegraphics[width=\columnwidth]{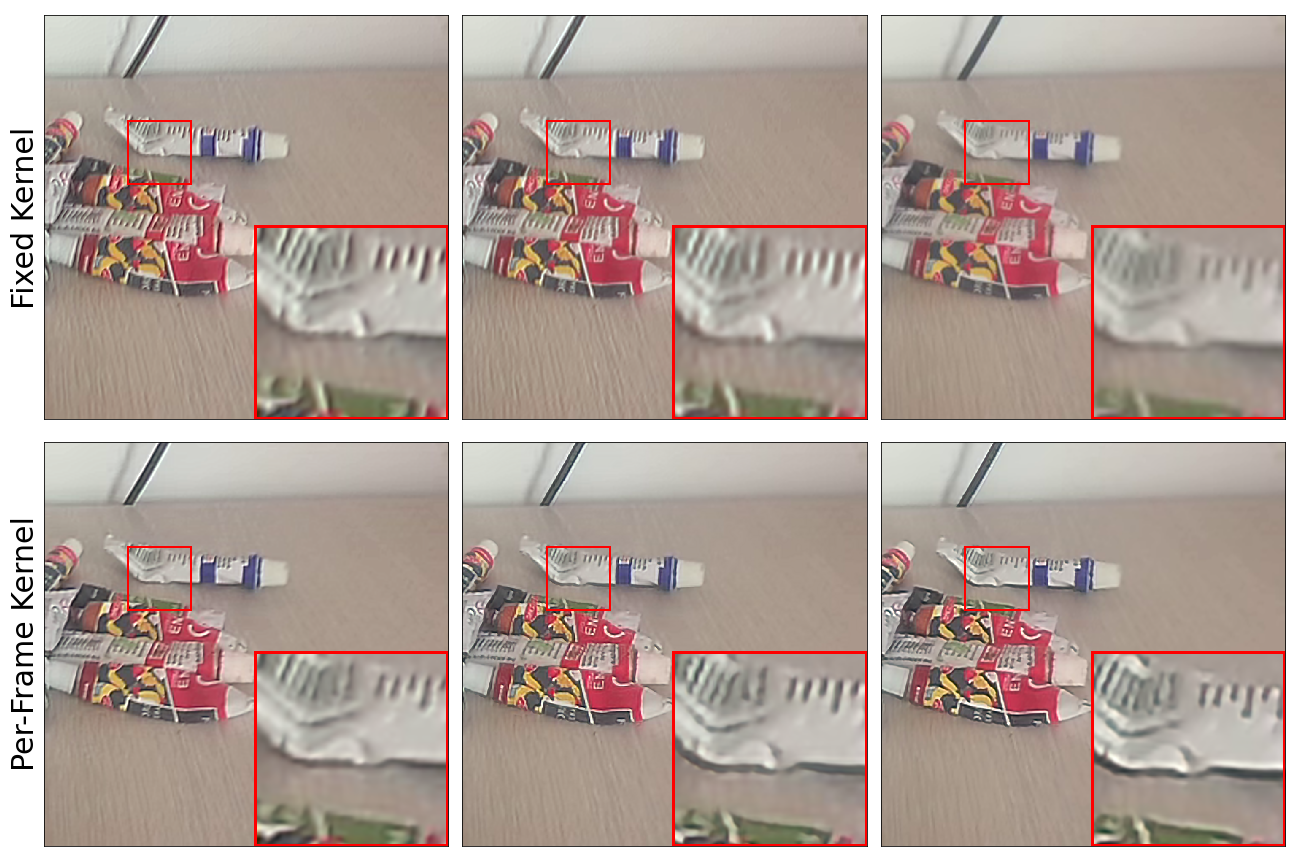}
 \caption{Additional Examples of consecutive frames in real-world videos taken from Something-Something~\cite{something} dataset upscaled using a fixed kernel (top in each example), and a different per-frame kernel (bottom in each example). Kernels are estimated using KernelGAN~\cite{KernelGAN} and the frames are restored using ZSSR~\cite{ZSSR}.}
 \label{sup_fig:teaser}
\end{figure*}

\begin{figure*}[!th]
 \centering
 \includegraphics[width=.9\textwidth ]{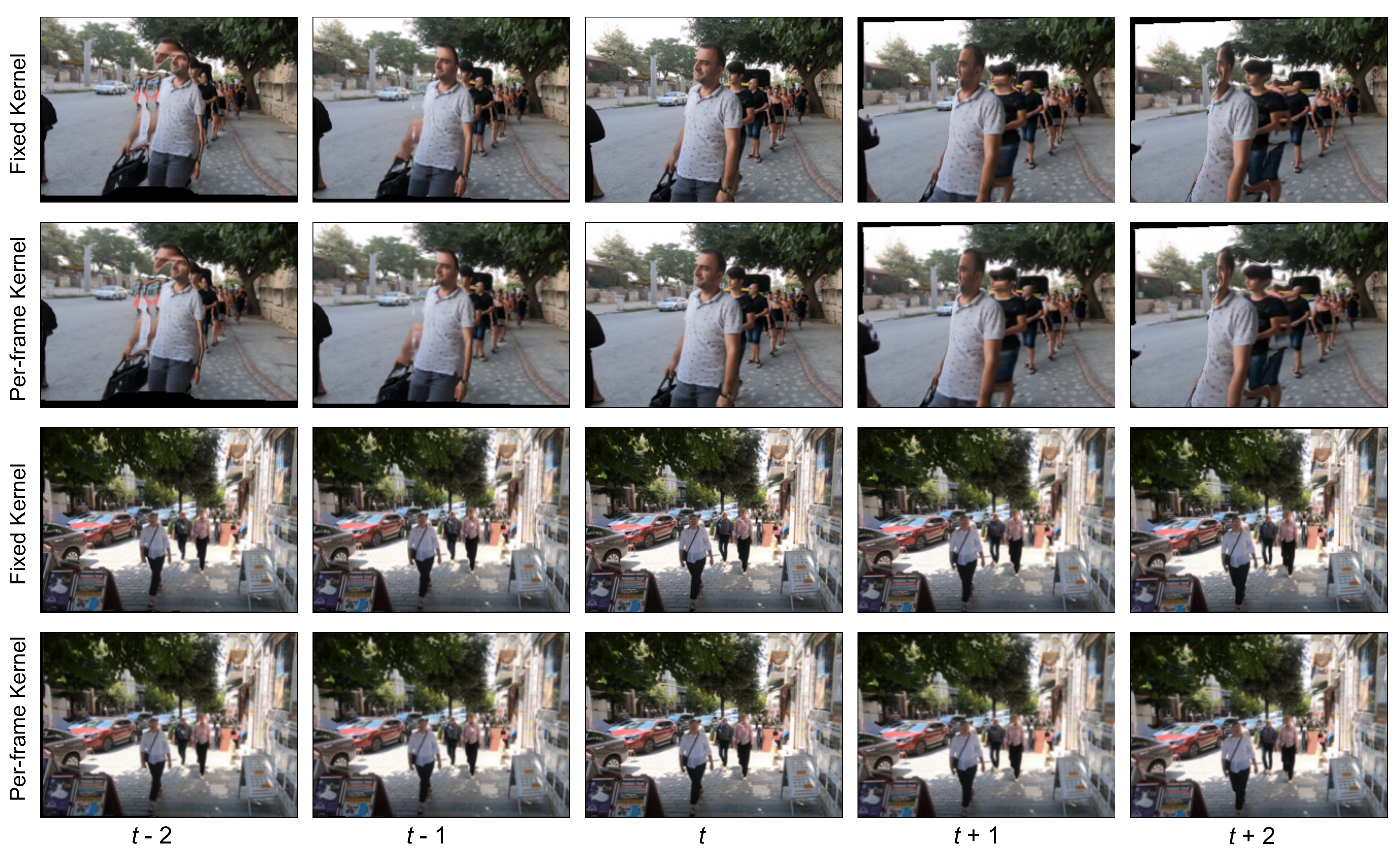}
 \caption{Additional example aligned frames at time step ${t-2, t-1, t+1, t+2}$ with their reference frame at time step $t$. The aligned frames are oversmoothed and blurred due to kernel mismatch for per-frame kernels found in real-world videos. In comparison, using a fixed downsampling kernel at every time step, which does not hold for real-world videos, leads to better motion compensation. Zoom in for best results.}
 \label{sup_fig:vid_flow}
\end{figure*}

\begin{figure*}[t!]
 \centering
 \includegraphics[width=\columnwidth]{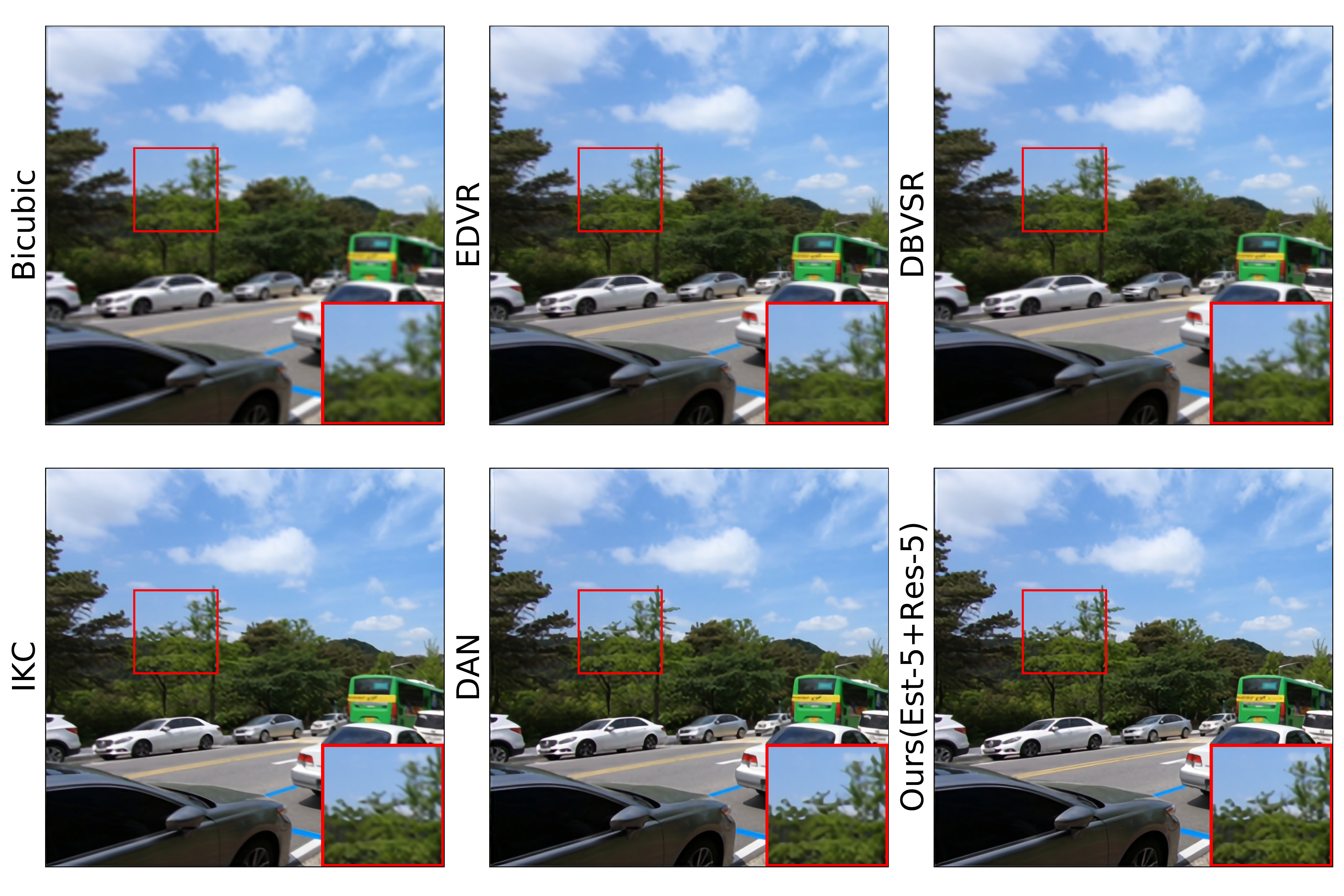}
 \includegraphics[width=\columnwidth]{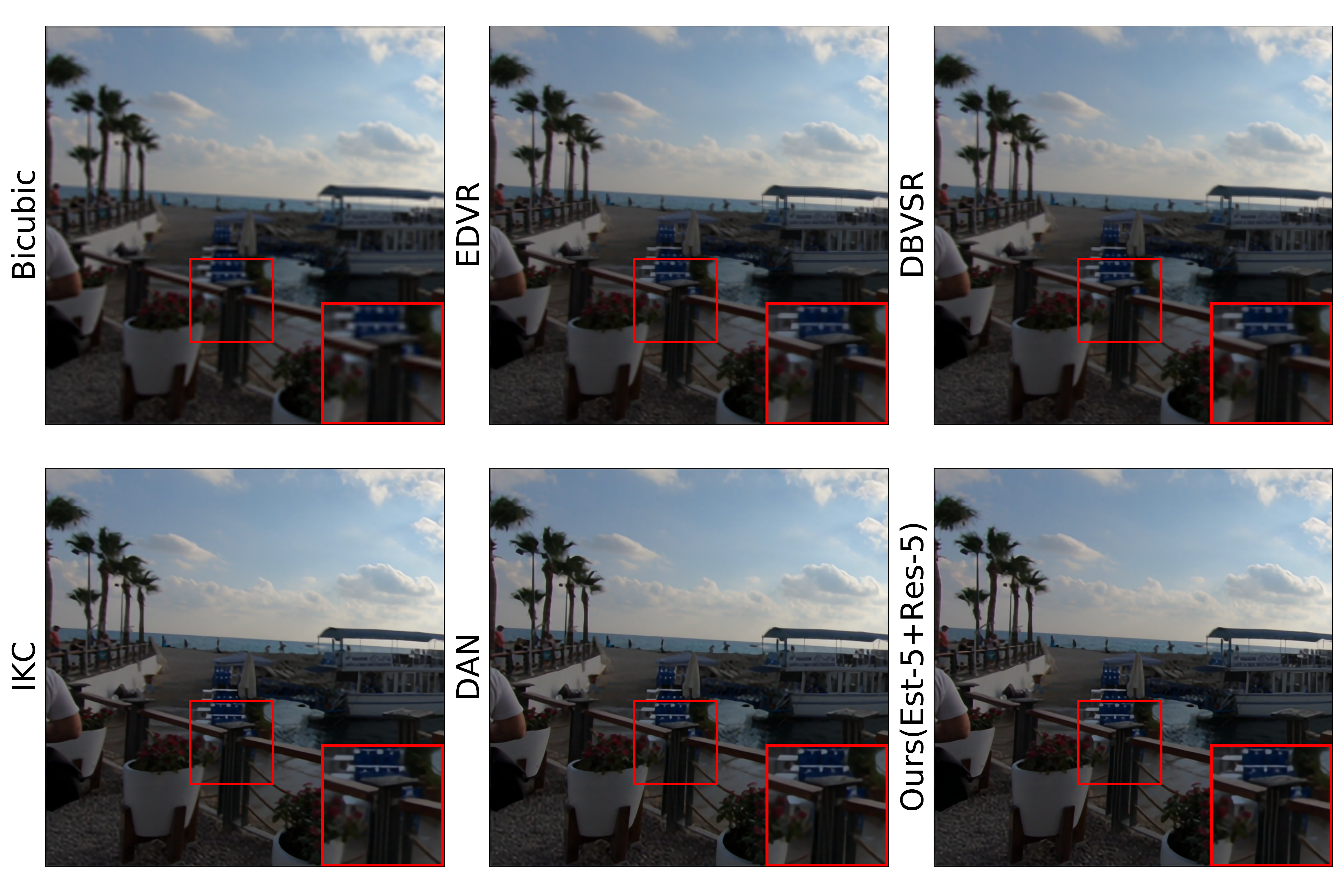}\\
 \includegraphics[width=\columnwidth]{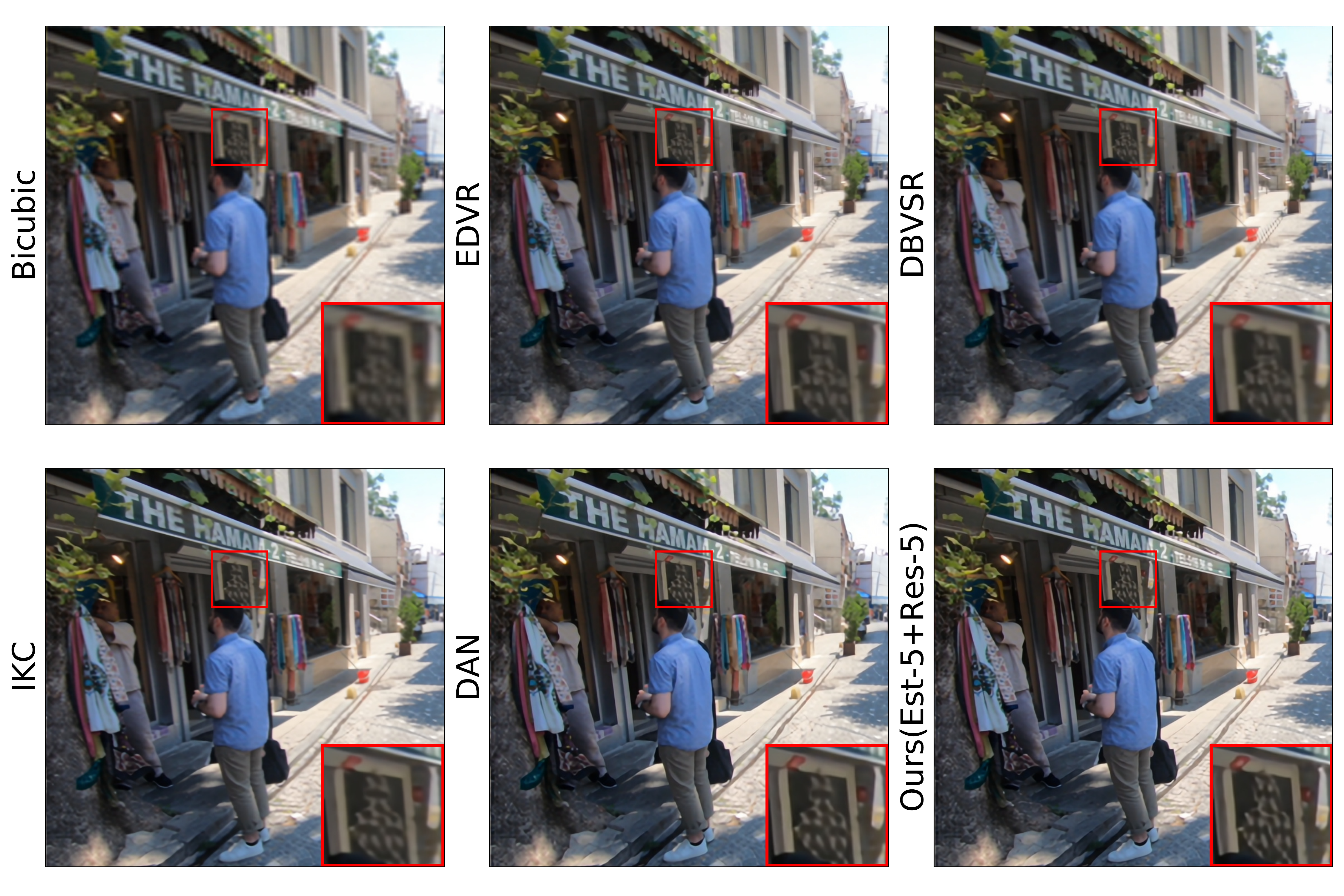} 
 \includegraphics[width=\columnwidth]{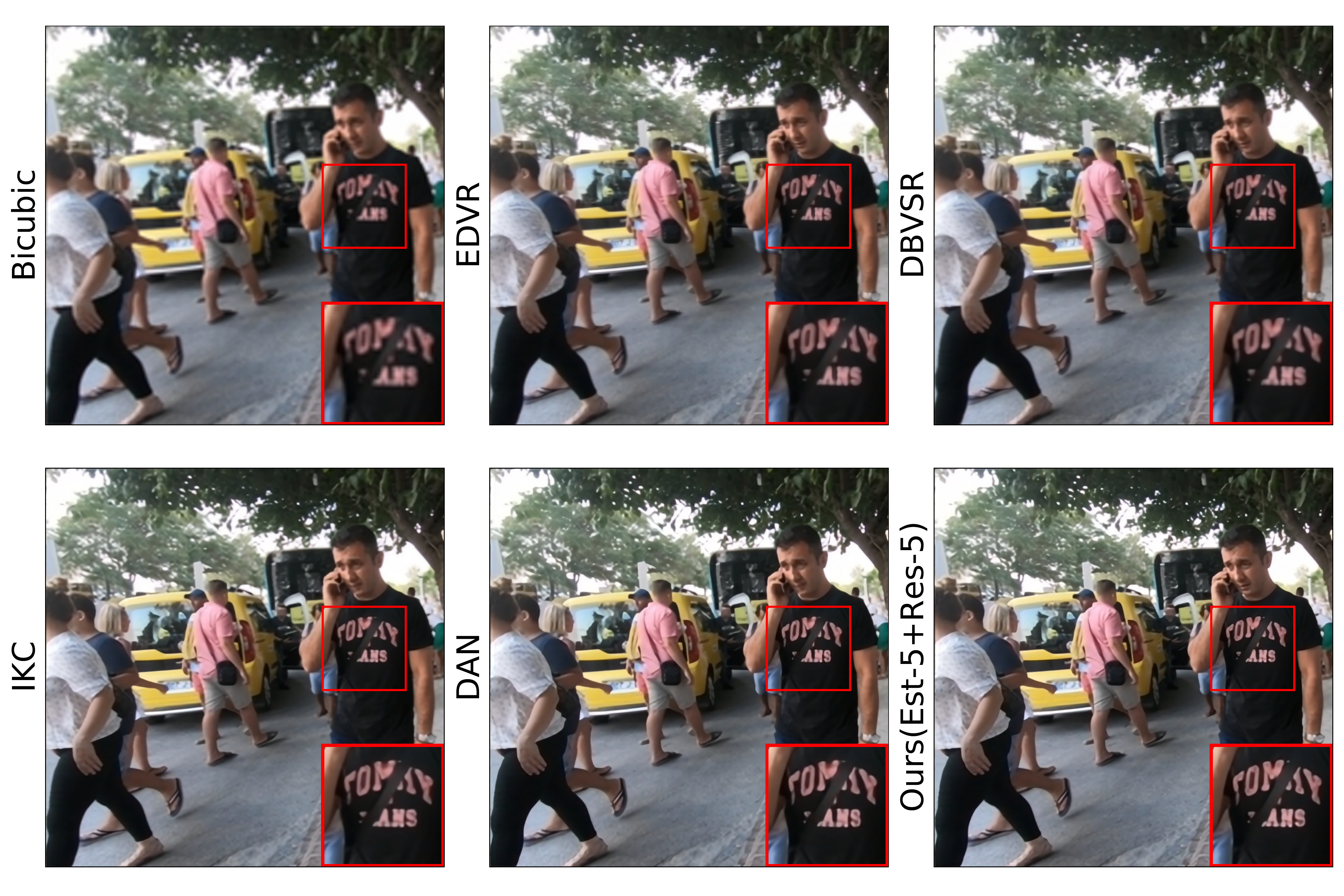}
 \caption{Qualitative comparison among existing models, along with bicubic upscaling, on our benchmark test sequences. Zoom in for best results.}
 \vspace{-5mm}
 \label{sup_fig:benchmarksm}
 
\end{figure*}

\begin{figure*}[ht!]
 \centering
 \includegraphics[width=\columnwidth]{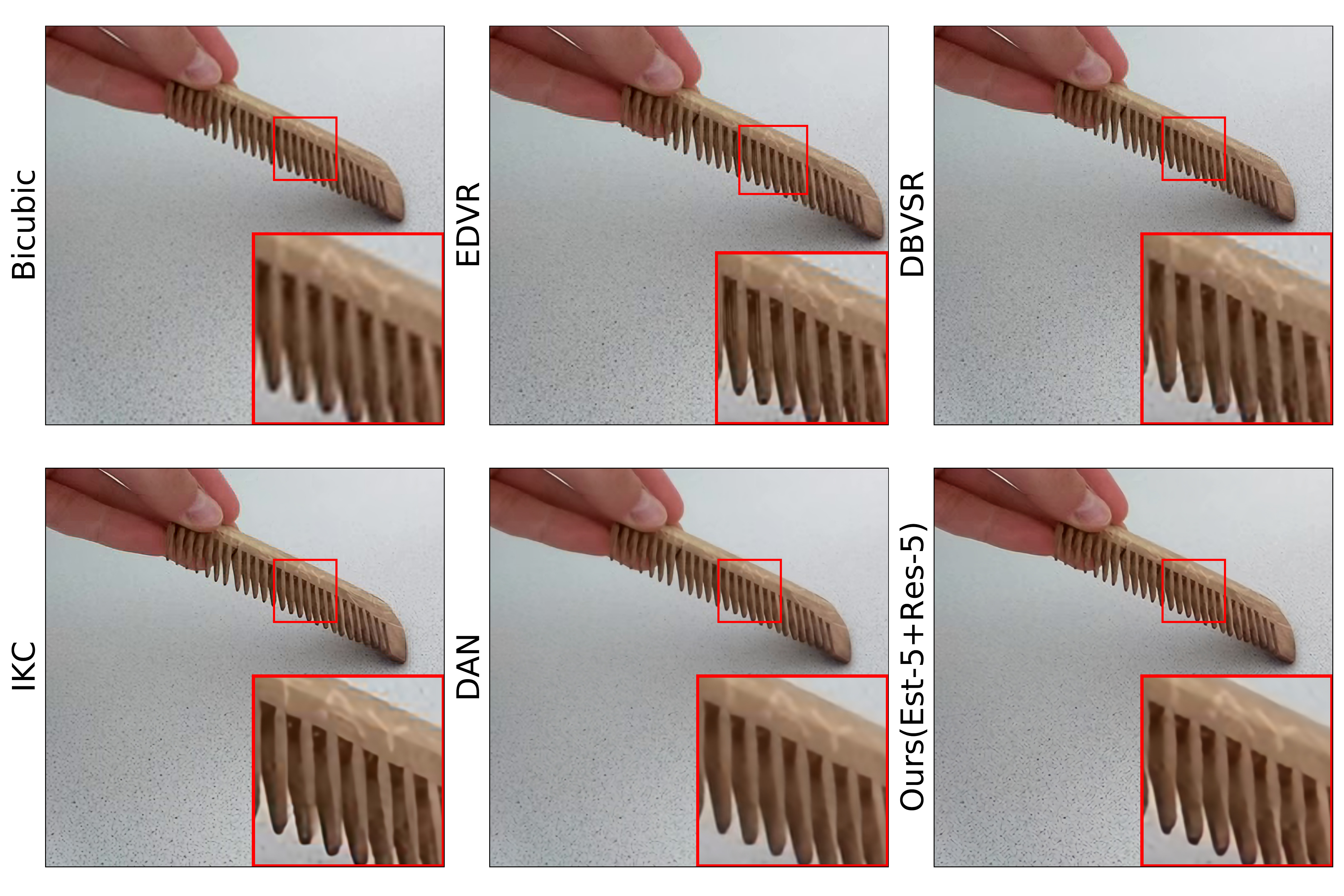}
 \includegraphics[width=\columnwidth ]{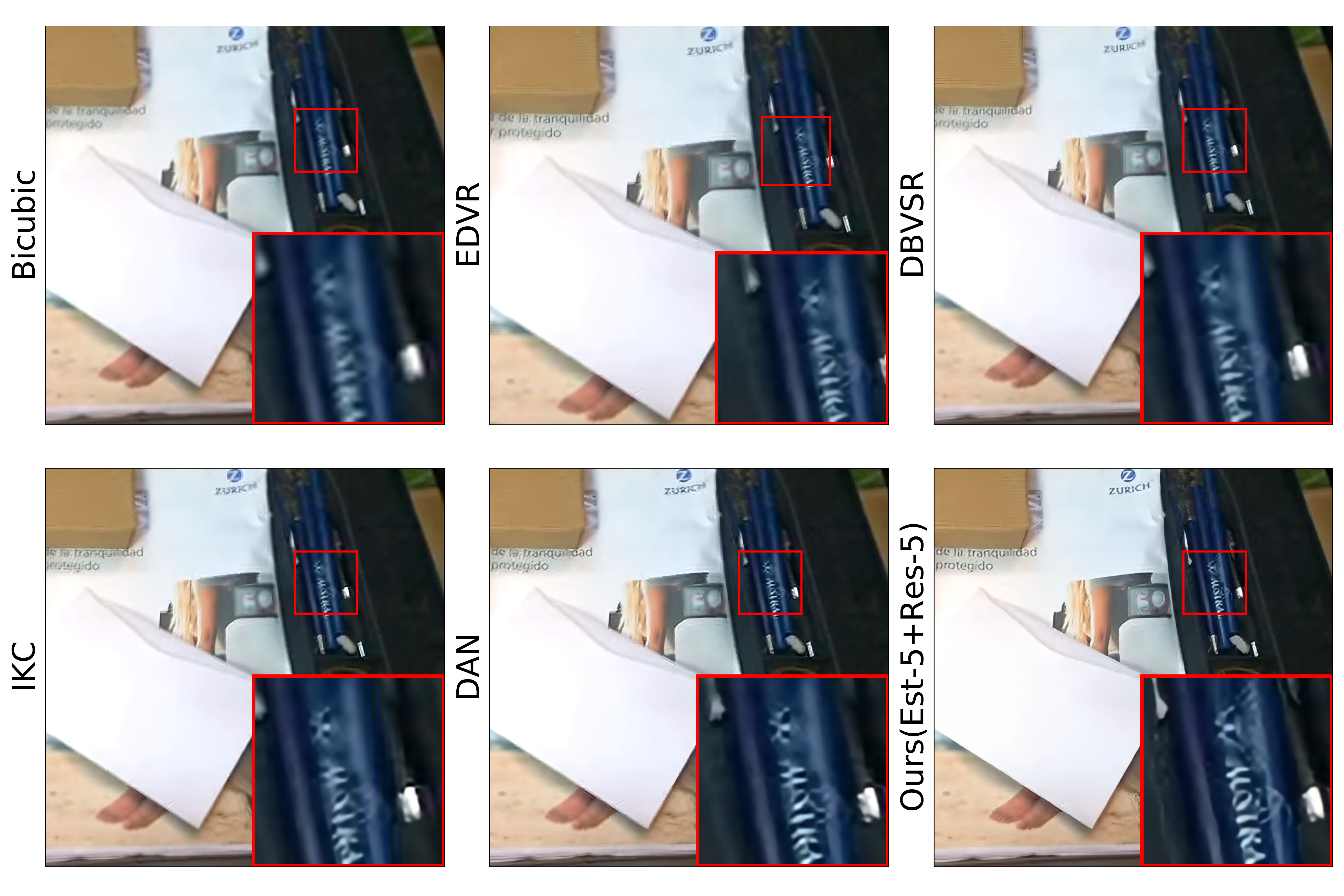}\\
 \includegraphics[width=\columnwidth]{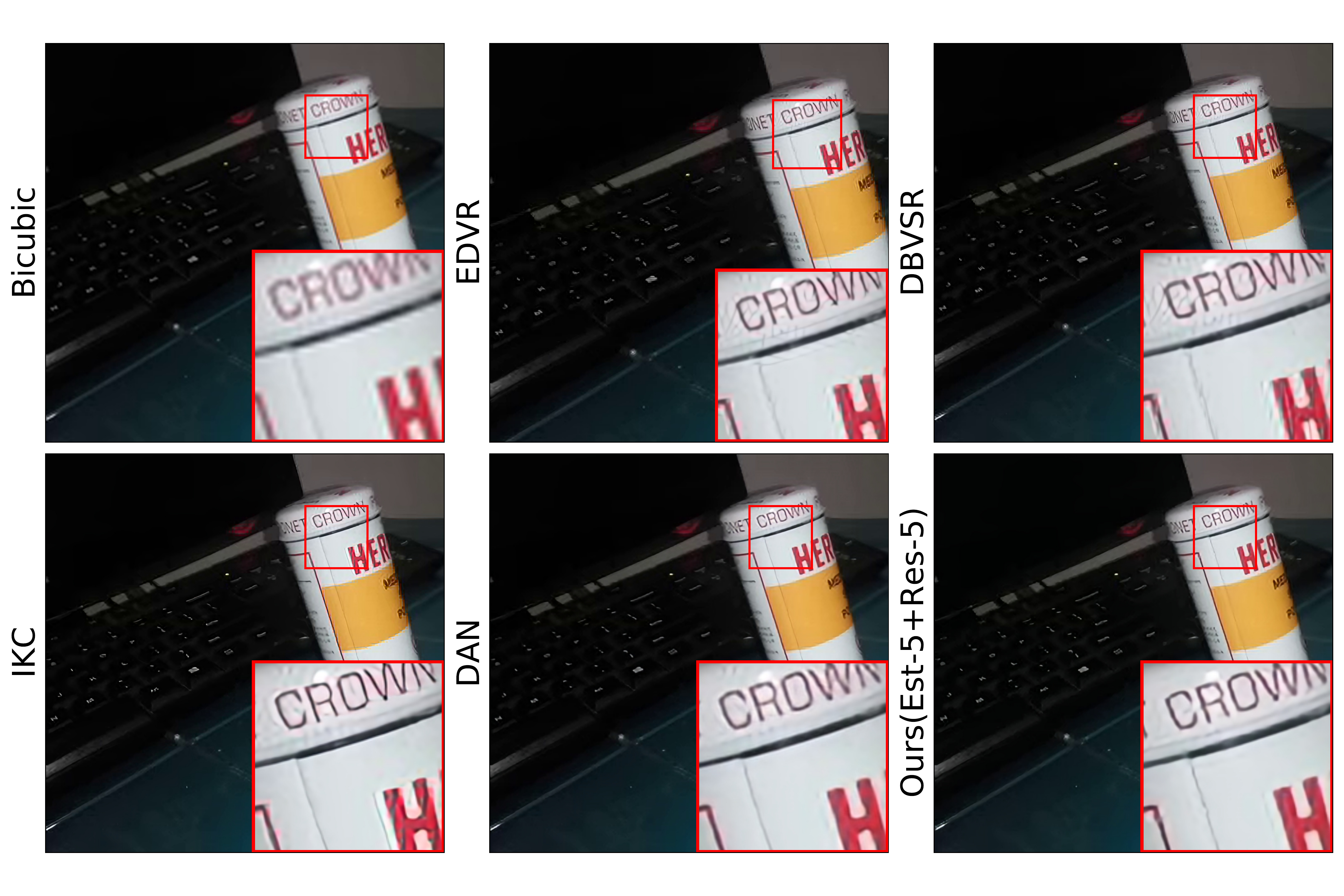}
   \includegraphics[width=\columnwidth]{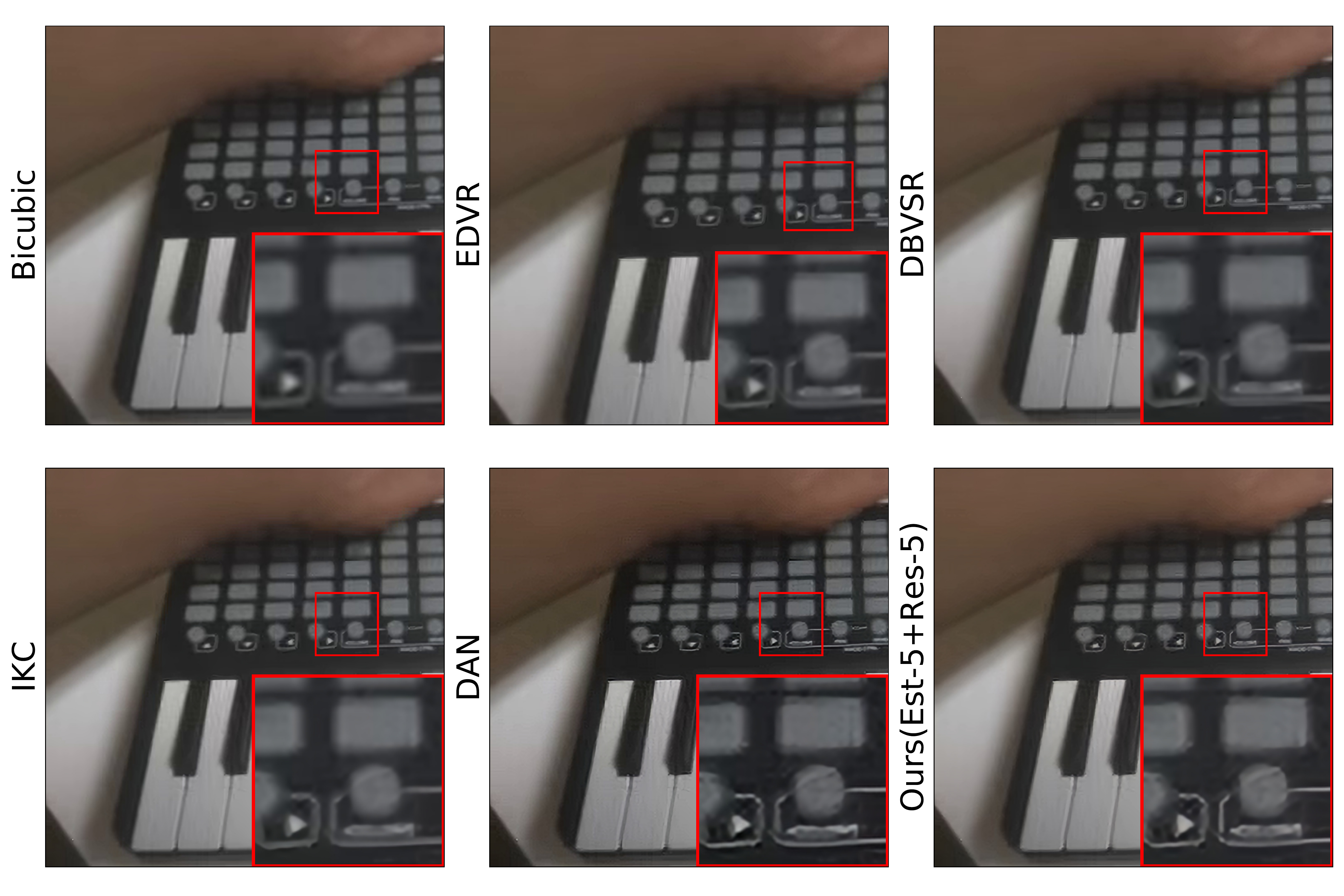}
 \caption{Real-world qualitative comparisons among existing models, along with bicubic upscaling. Zoom in for best results. Note that there is no ground-truth available.}
 \vspace{-2mm}
 \label{sup_fig:realworldsm}
\end{figure*}

We provided additional results here due to space limitations in the main paper.
Fig.~\ref{sup_fig:teaser} provides additional examples for Fig.~3 of the main paper, highlighting that using a fixed kernel to upscale all the frames in a video can result in inferior restoration outcomes as compared to using a per-frame kernel even without incorporating temporal frame information.
Likewise, Fig.~\ref{sup_fig:vid_flow} shows the additional examples for Fig.~4 of the main paper, underlining that the use of explicit motion compensation in previous works for temporal alignment results in more errors when applied to real-world videos.
Fig.~\ref{sup_fig:benchmarksm} and Fig.~\ref{sup_fig:realworldsm} provides additional qualitative examples comparing our multi-frame SR model with previous multi-frame SR and blind image-based SR models on \textit{REDS10} and real-world videos respectively.
Lastly, we provided a sample video along with this document.


\end{document}